\documentclass[journal]{IEEEtran}
\usepackage{cite}
\usepackage{graphicx}
\usepackage{amssymb,amsmath,bm}
\usepackage[colorlinks,linkcolor=black,anchorcolor=black,citecolor=black,urlcolor=black]{hyperref}
\usepackage{hyperref}
\usepackage{multirow}
\usepackage{tabularx}
\ifCLASSINFOpdf
\else
\fi
\begin{document}
%
\title{A Neural Vocoder with Hierarchical Generation of Amplitude and Phase Spectra for Statistical Parametric Speech Synthesis}
%
%
%

\author{Yang~Ai,~\IEEEmembership{Student Member,~IEEE},~Zhen-Hua~Ling,~\IEEEmembership{Senior Member,~IEEE}
\thanks{This work was partially funded by  the National Natural Science Foundation of China (Grants No. 61871358).}
}

%
%

\markboth{Journal of \LaTeX\ Class Files,~Vol.~6, No.~1, January~2007}%
{Shell \MakeLowercase{\textit{et al.}}: Bare Demo of IEEEtran.cls for Journals}
%



\maketitle

\begin{abstract}
This paper presents a neural vocoder named HiNet which reconstructs speech waveforms from acoustic features
by predicting amplitude and phase spectra hierarchically.
Different from existing neural vocoders such as WaveNet, SampleRNN and WaveRNN which directly generate waveform samples using single neural networks,
the HiNet vocoder 
is composed of an amplitude spectrum predictor (ASP) and a phase spectrum predictor (PSP).
The ASP is a simple DNN model which predicts log amplitude spectra (LAS) from acoustic features.
The predicted LAS are sent into the PSP for phase recovery.
Considering the issue of phase warping and the difficulty of phase modeling,
the PSP is constructed by concatenating a neural source-filter (NSF) waveform generator with a phase extractor.
We also introduce generative adversarial networks (GANs) into both ASP and PSP.
Finally, the outputs of ASP and PSP are combined to reconstruct speech waveforms by short-time Fourier synthesis.
Since there are no autoregressive structures in both predictors, the HiNet vocoder can generate speech waveforms with high efficiency.
Objective and subjective experimental results show that our proposed HiNet vocoder achieves better naturalness of reconstructed speech than the conventional STRAIGHT vocoder, a 16-bit WaveNet vocoder using open source implementation and an NSF vocoder with similar complexity to the PSP and obtains similar performance with a 16-bit WaveRNN vocoder.
We also find that the performance of HiNet is insensitive to the complexity of the neural waveform generator in PSP to some extend.
After simplifying its model structure, the time consumed for generating 1s waveforms of 16kHz speech using a GPU can be further reduced from 0.34s to 0.19s without significant quality degradation. 

\end{abstract}

\begin{IEEEkeywords}
vocoder, neural network, amplitude spectrum, phase spectrum, statistical parametric speech synthesis
\end{IEEEkeywords}

%
\IEEEpeerreviewmaketitle

\section{Introduction}
\label{sec1: Introduction}
\IEEEPARstart{S}
{peech} synthesis, a technology that converts texts into speech waveforms, plays a more and more important role in people's daily life.
A speech synthesis system with high intelligibility, naturalness and expressiveness is a goal pursued by speech synthesis researchers.
Recently, statistical parametric speech synthesis (SPSS) has become a widely used speech synthesis framework due to its flexibility achieved by acoustic modeling and vocoder-based waveform generation.
Hidden Markov models (HMMs) \cite{tokuda2013speech}, deep neural networks (DNNs) \cite{ze2013statistical}, recurrent neural networks (RNNs) \cite{fan2014tts} and other deep learning models \cite{ling2015deep} have been applied to build the acoustic models for SPSS.
Vocoders \cite{dudley1939vocoder} which reconstruct speech waveforms from acoustic features (e.g., mel-cepstra and F0) also play an important role in SPSS. Their performance affects the quality of synthetic speech significantly.
Some conventional vocoders, such as STRAIGHT \cite{kawahara1999restructuring} and WORLD \cite{morise2016world} which are designed based on the source-filter model of speech production \cite{gunnar1960acoustic}, have been popularly applied in current SPSS systems.
However, these vocoders still have some deficiencies, such as the loss of spectral details and phase information.

Recently, some neural generative models for raw audio signals \cite{oord2016wavenet,mehri2016samplernn,kalchbrenner2018efficient} have been proposed and demonstrated good performance.
For example, WaveNet \cite{oord2016wavenet} and SampleRNN \cite{mehri2016samplernn} predicted the distribution of each waveform sample conditioned on previous samples and additional conditions using convolutional neural networks (CNNs) and RNNs respectively.
These models represented waveform samples as discrete symbols. Although the $\mu$-law quantization strategy \cite{recommendation1988g} has been applied, the neural waveform generators with low quantization bits (e.g., 8-bit or 10-bit) always suffered from perceptible quantization errors.
In order to achieve 16-bit quantization of speech waveforms, the WaveRNN model \cite{kalchbrenner2018efficient} was proposed, which generated 16-bit waveforms by splitting the RNN state into two parts and predicting the 8 coarse bits and the 8 fine bits respectively.
However, due to the autoregressive generation manner, these models were very inefficient at generation stage.
Therefore, some variants such as knowledge-distilling-based models (e.g., parallel WaveNet \cite{oord2017parallel} and ClariNet \cite{ping2018clarinet}) and flow-based models (e.g., WaveGlow \cite{prenger2018waveglow}) were then proposed to improve the efficiency of generation.

\begin{figure*}[t]
    \centering
    \includegraphics[height=5cm]{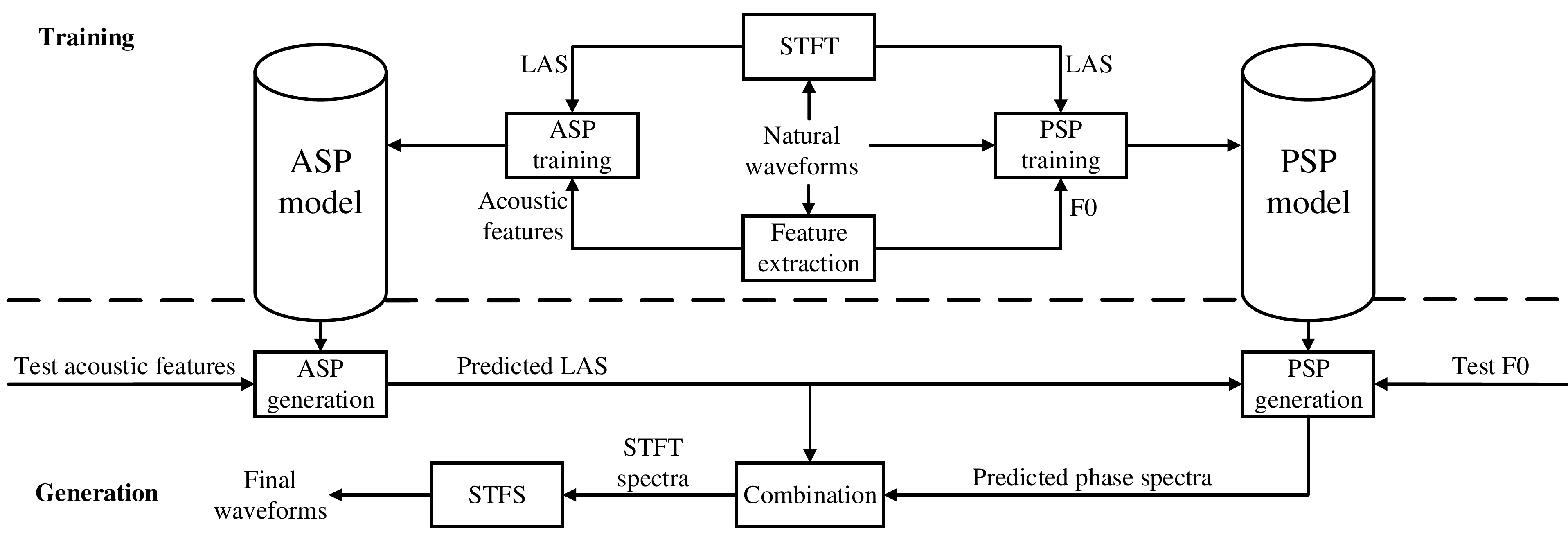}
    \caption{The flowchart of the training and generation processes of our proposed HiNet vocoder. Here, \emph{ASP}, \emph{PSP} and \emph{LAS} stand for amplitude spectrum predictor, phase spectrum predictor and log amplitude spectra respectively.}
    \label{fig: flowchart}
\end{figure*}

Neural vocoders  based on these waveform generation models \cite{tamamori2017speaker,hayashi2017investigation,adiga2018use,ai2018samplernn,ai2019dnn,lorenzo2018robust}
have been developed to reconstruct speech waveforms from various acoustic features for SPSS and some other tasks,
such as voice conversion \cite{liu2018wavenet,kobayashi2017statistical}, bandwidth extension \cite{ling2018waveform} and speech coding \cite{klejsa2018high}.
Experimental results confirmed that these neural vocoders performed significantly better than conventional ones.
Some improved neural vocoders, such as glottal neural vocoder \cite{cui2018new,juvela2018speaker,juvela2019glotnet}, LP-WaveNet \cite{hwang2018lp}, LPCNet \cite{valin2019lpcnet}, and neural source-filter  (NSF) vocoder \cite{wang2019neural}, have been further proposed by combining speech production mechanisms with
neural networks and have also demonstrated impressive performance.
The first three vocoders predict the excitation waveforms by a neural network and then the excitation waveforms pass through a vocal tract filter to generate the final waveforms, which imitate the process of linear prediction (LP).
The last one achieves the process of the source-filter model by neural networks rather than the conventional models.
However, none of these models consider recovering waveforms from separately predicted amplitude and phase spectra.

There are still some limitations with current neural vocoders and the most significant one is that they have much higher computation complexity than conventional STRAIGHT and WORLD vocoders.
The autoregressive neural vocoders (e.g., WaveNet, SampleRNN and WaveRNN) are very inefficient at synthesis time due to their point-by-point generation process.
For knowledge-distilling-based vocoders (e.g., parallel WaveNet and ClariNet), although the student model accelerates the generation process by removing autoregressive connections, they require a WaveNet as the teacher model to guide the training of the student model and additional criteria such as power loss and spectral amplitude distance must be used.
These facts make it difficult to train knowledge-distilling-based vocoders.
The flow-based vocoders (e.g., WaveGlow) are also efficient due to the flow-based model without any autoregressive connections.
However, the complexity of model structures of WaveGlow \cite{prenger2018waveglow} is reported to be huge with low training efficiency.

This paper explores the approaches to improve the run-time efficiency of neural vocoders by combining neural waveform generation models with the frequency-domain representation of speech waveforms.
Inspired by the knowledge that speech waveforms can be perfectly reconstructed from their short-time Fourier transform (STFT) results
which consist of frame-level amplitude spectra and phase spectra,
this paper proposes a neural vocoder which recovers speech waveforms by predicting amplitude and phase spectra hierarchically from input acoustic features.
We name this vocoder \emph{HiNet} because it is expected to generate waveforms with \emph{hi}gh quality and \emph{hi}gh efficiency by \emph{hi}erarchical prediction.
Different from existing neural vocoders which directly generate waveform samples using single neural networks,
the HiNet vocoder 
is composed of an amplitude spectrum predictor (ASP) and a phase spectrum predictor (PSP).
The ASP is a simple DNN which predicts frame-level log amplitude spectra (LAS) from acoustic features.  Then, the predicted LAS are sent into the PSP for phase recovery.
Considering the issue of phase warping and the difficulty of phase modeling,
the PSP is constructed by concatenating a neural waveform generator with a phase extractor.
Since the task of the neural waveform generator in PSP is not to generate the final waveforms but to supplement the amplitude spectra with phase information,
some light-weight models can be adopted even if their overall prediction accuracy is not perfect.
In our implementation, the neural waveform generator is built by adapting the non-autoregressive NSF vocoder  \cite{wang2019neural} from three aspects.
First, LAS are used as the input of PSP rather than spectral features (e.g., mel-cepstra).
Second, the initial phase of the sine-based excitation signal is pre-calculated  for each voiced segment at the training stage of the PSP to benefit phase modeling.
Third, 
a waveform loss and a correlation loss are introduced into the complete loss function in order to enhance its ability of measuring phase distortion.
Besides, generative adversarial networks (GANs) \cite{goodfellow2014generative} are also introduced into ASP and PSP to fit the true distribution of amplitude and phase spectra.
Finally, the outputs of ASP and PSP are combined to recover speech waveforms by short-time Fourier synthesis (STFS).
Experimental results show  that the proposed HiNet vocoder achieves better naturalness of reconstructed speech than the conventional STRAIGHT vocoder, a 16-bit WaveNet
vocoder implemented by public source codes and an NSF vocoder with similar complexity to the PSP, and obtains similar performance with a 16-bit WaveRNN vocoder.

\begin{figure*}[t]
    \centering
    \includegraphics[height=2.45cm]{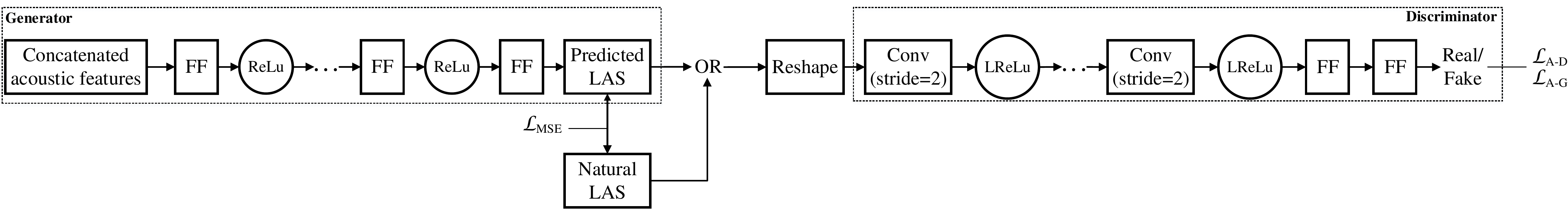}
    \caption{Model structure of the amplitude spectrum predictor (ASP). Here, \emph{FF} and \emph{Conv} represent feed-forward and convolutional layers (stride size=2) respectively, \emph{ReLu} and \emph{LReLu} represent rectified linear units and leaky rectified linear units respectively.}
    \label{fig: ASP}
\end{figure*}

There are two main characteristics of the HiNet vocoder.
First, there are no autoregressive structures in both predictors. Thus, the HiNet vocoder is able to generate speech waveforms with high efficiency by parallel computation.
Second, the neural waveform generator only contributes to the prediction of phase spectra.
Further experimental results reveal that the performance of HiNet is insensitive to the complexity of the neural waveform generator in PSP to some extend.
After simplifying its model structure, the training efficiency is improved and the time consumed for generating 1s waveforms of 16kHz speech can be further reduced from 0.34s to 0.19s without significant quality degradation.



This paper is organized as follows. In Section \ref{sec2: Proposed Method}, we present our proposed HiNet vocoder in detail.
Section \ref{sec3: Experiments} reports our experimental results and conclusions are given in Section \ref{sec4: Conclusion}.

\section{Proposed Methods}
\label{sec2: Proposed Method}
The proposed HiNet vocoder consists of an amplitude spectrum predictor (ASP) and a phase spectrum predictor (PSP).
The flowchart of its training and generation processes is illustrated in Fig. \ref{fig: flowchart}.

\subsection{Amplitude Spectrum Predictor}
\label{subsec2A: Amplitude Spectrum Predictor}

The ASP predicts LAS from input acoustic features which include mel-cepstra, energy, F0 and voiced/unvoicded (V/UV) flag.
For better generation efficiency, a simple DNN consisting of multiple FF layers without any autoregressive structures is adopted to build the ASP as shown in the generator (denoted by $G_A$) in Fig. \ref{fig: ASP}.

Let $\bm{a}_n=[a_{n,1},\dots,a_{n,C}]^\top$ and $\bm{l}_n=[l_{n,1},\dots,l_{n,K}]^\top$ denote the acoustic features and the LAS at the $n$-th frame respectively,
where $n$, $c$ and $k$ represent the frame index, the dimension index and the frequency bin index, $C$ and $K$ denote the total numbers of acoustic feature dimensions and frequency bins respectively.
For utilizing history information, the model input is a concatenation of current frame and $n_p$ previous frames (i.e., $\bm{a}_n^C=[\bm{a}_{n-n_p}^\top,\dots,\bm{a}_{n-1}^\top,\bm{a}_n^\top]^\top$).
The model output is the LAS of current frame as shown in Fig. \ref{fig: ASP}.

If GAN is not used in the ASP, only the generator in Fig. \ref{fig: ASP} is trained.
At the training stage, parallel concatenated acoustic features $\bm{a}_{1:N}^C$ (denoted by $\bm{a}^C$) and LAS $\bm{l}_{1:N}$ (denoted by $\bm{l}$) are extracted from natural waveforms and $N$ is the total number of frames.
The training criterion is to minimize the mean square error (MSE) between the predicted LAS $\hat{\bm{l}}=G_A(\bm{a}^C)$ and the real LAS $\bm{l}$ as
\begin{align}
\label{equ: MSE on LAS}
\mathcal L_{MSE}=\dfrac{1}{NK}\sum_{n}\sum_{k}(l_{n,k}-\hat{l}_{n,k})^2+L_2\_reg,
\end{align}
where $\hat{\bm{l}}_n=[\hat{l}_{n,1},\dots,\hat{l}_{n,K}]$ is the predicted LAS at the $n$-th frame.
$L_2\_reg$ is an L2 regularization term of all weights in the model for avoiding overfitting.

At the generation stage, a global mean normalization (GMN) operation is conducted 
to compensate the global distortion between the amplitude spectra predicted by the DNN and the natural ones.
For the $k$-th frequency bin, a compensation factor $q_k$ is estimated given the trained DNN as 
\begin{align}
\label{equ: compensation factor}
q_k=\dfrac{\sum_{n}\exp(l_{n,k})}{\sum_{n}\exp(\hat{l}_{n,k})},
\end{align}
where $n$ denotes the frame index of the training set.
The vector $\bm{q}=[q_1,\dots,q_K]^\top$  further passes through a median filter along the frequency axis to get a smoothed curve $\bm{q}^{mf}=[q_1^{mf},\dots,q_K^{mf}]^\top$.
The final LAS at each frame is obtained by
\begin{align}
\label{equ: amplitude enhancement}
\hat{\bm{l}}_n^{FNL}=\log\left(\exp(\hat{\bm{l}}_n)\odot\bm{q}^{mf}\right),
\end{align}
where $\odot$ represents element-wise product.

If GAN is used in the ASP, a discriminator (denoted by $D_A$) as shown in Fig \ref{fig: ASP} is also trained along with the generator.
The discriminator consists of multiple convolutional layers which operate along with the frequency axis of the input LAS. The input LAS of the discriminator which is obtained by reshaping the predicted LAS or natural LAS has width $K$ and one channel.
Leaky rectified linear units \cite{maas2013rectifier} are used as the activation function.
Each convolutional block downsamples the input LAS by a factor of two, using strided convolutions, until a small width is reached.
The amount of filters per convolutional layer increases so that the channel gets larger as the width gets narrower.
Finally, two feed-forward layers both with one neuron are used to reduce the channel and width to 1 respectively and the results are used to define loss functions for GAN.

\begin{figure*}
    \centering
    \includegraphics[height=4.25cm]{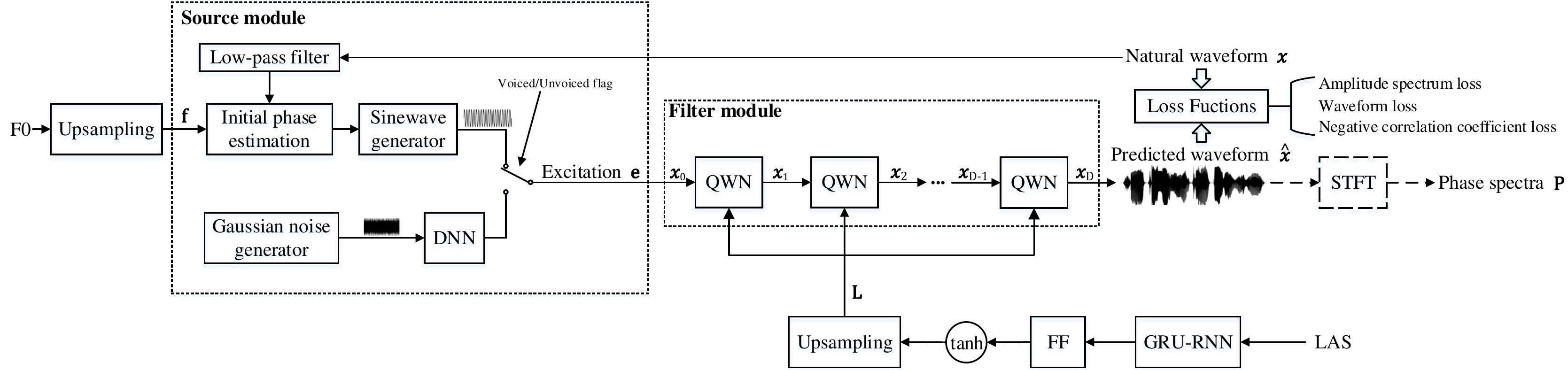}
    \caption{Model structure of the phase spectrum predictor (PSP). Here, \emph{FF} and \emph{GRU-RNN} represent feed-forward and unidirectional GRU-based recurrent layers respectively, \emph{DNN} represents a deep neural network with multiple FF layers, \emph{tanh} denotes the hyperbolic tangent functions, \emph{QWN} is a quasi WaveNet structure as shown in Fig. \ref{fig: QWN}. The dotted lines indicates the source module and filter module. The dashed lines represent the operations only used at the generation stage.}
    \label{fig: PSP}
\end{figure*}

At the training stage, a Wasserstein GAN \cite{gulrajani2017improved } loss is used for training.
To keep the discriminator Lipschitz continuous, a gradient penalty \cite{mescheder2018training} is also introduced into the loss function of the discriminator.
Therefore, the loss function of the discriminator is defined as
\begin{align}
\label{equ: ASP-D-loss}
\begin{split}
\mathcal L_{A-D}=&-\mathbb{E}_{\bm{l}\sim P_r}[D_A(\bm{l})]+\mathbb{E}_{\hat{\bm{l}}\sim P_g}[D_A(\hat{\bm{l}})]\\
&+\lambda_{A-GP}\cdot\mathbb{E}_{\bm{l}\sim P_r,\hat{\bm{l}}\sim P_g}[(\max\{0,\Vert\nabla_{\tilde{\bm{l}}}D_A(\tilde{\bm{l}})\Vert-1\})^2],
\end{split}
\end{align}
where $\tilde{\bm{l}}=\varepsilon \bm{l}+(1-\varepsilon)\hat{\bm{l}}$ is sampled randomly along the line segment between $\bm{l}$ and $\hat{\bm{l}}$ and $\lambda_{A-GP}$ is a hyperparameter.
Regard with the generator, in addition to GAN-related loss, MSE loss $\mathcal L_{MSE}$ is also used as an auxiliary loss.
Therefore, the loss function of the generator is defined as
\begin{align}
\label{equ: ASP-G-loss}
\mathcal L_{A-G}=-\mathbb{E}_{\hat{\bm{l}}\sim P_g}[D_A(\hat{\bm{l}})]+\lambda_{MSE}\cdot\mathcal L_{MSE},
\end{align}
where $\lambda_{MSE}$ is a hyperparameter. The training process is divided into three steps: first using $\mathcal L_{MSE}$ to train the generator, then using $\mathcal L_{A-D}$ to train the discriminator, and finally using $\mathcal L_{A-G}$ and $\mathcal L_{A-D}$ to train the generator and discriminator alternately like a standard GAN training process \cite{goodfellow2014generative}.
GMN is not used at the generation stage because GAN is expected to compensate the global distortion on amplitude spectra and effectively alleviate the over-smoothing problem when only using the MSE loss to train the model.

We can see that the whole ASP model operates at frame-level without any autoregressive calculations. Therefore, its training and generation processes are very efficient.

\subsection{Phase Spectrum Predictor}
\label{subsec2B: Phase Spectrum Predictor}
The aim of PSP is to recover phase spectra given input amplitude spectra.
However,  modeling and predicting phase spectra directly are difficult due to the issue of phase warping.
Since temporal waveforms contain the information of both amplitude and phase spectra,
this paper proposes to predict phase spectra utilizing a neural waveform generator.
For predicting phase spectra efficiently, the waveform generator is designed based on the non-autoregressive NSF vocoder \cite{wang2019neural}.
Several modifications are made in order to focus on recovering  phase spectra from input amplitude spectra.
Finally, the phase spectra extracted from the waveforms generated by the neural waveform generator are used as the outputs of PSP.

The structure of PSP is illustrated in Fig. \ref{fig: PSP},
which first converts input LAS and F0 sequences into waveforms $\hat{\bm{x}}=[\hat{x}_1,\dots,\hat{x}_T]^\top$ using a neural waveform generator and
then extracts phase spectra $\bm{P}=[\bm{p}_1,\dots,\bm{p}_N]$ from $\hat{\bm{x}}$ by STFT analysis.
At the training stage, LAS and F0 sequences are calculated from  natural waveform ${\bm{x}}=[{x}_1,\dots,{x}_T]^\top$
and the loss functions are defined between $\bm{x}$ and $\hat{\bm{x}}$.
At the generation stage, the PSP adopts the test F0 sequence and the LAS predicted by ASP as inputs. 
Similar to the NSF vocoder \cite{wang2019neural}, the neural waveform generator in PSP consists of a source module and a filter module.
The details of these two modules and the loss functions will be introduced in the following subsections.

\subsubsection{Source Module}
\label{subsec2B1: Source Module}
The upsampled F0 sequence $\bm{f}=[f_1,\dots,f_T]^\top$ is obtained by repeating the F0 values within each frame, and is used as the input of the source module.
The output of the source module is an excitation signal $\bm{e}=[e_1,\dots,e_T]^\top$, which is a sine-based signal for voiced segments and a DNN-transformed Gaussian white noise for unvoiced segments.
Mathematically, for time step $t$, the excitation signal $e_t$ is defined as
\begin{align}
\label{equ: excitation signal}
e_t=\left\{\begin{array}{ll}\alpha\sin(\sum\limits_{h=1}^t2\pi f_h\dfrac{1}{N_s}+\phi_j)+n_t,&f_t>0,t\in V_j\\ g(\dfrac{1}{3\sigma}n_t),&f_t=0\end{array}\right.,
\end{align}
where $f_t=0$ denotes that the $t$-th sampling point belongs to an unvoiced frame,
 $g(\cdot)$ represents a DNN-based transformation, $n_t\sim\mathcal N(0,\sigma^2)$ is a Gaussian white noise at time $t$, $N_s$ is the sampling rate of waveforms,
$V_j$ is the $j$-th voiced segment that the $t$-th sampling point belongs to,
$\phi_j\in(-\pi,\pi]$ is the initial phase of the $j$-th voiced segment, 
$\alpha$ and $\sigma$ are hyperparameters.
At the training stage, we estimate the initial phase $\phi_j$ of each voiced segment for better phase modeling.
First, the $j$-th voiced segment of the natural waveform $\bm{x}$ passes through a low-pass filter whose cut-off frequency is the maximal F0 of this segment
in order to obtain a reference waveform without formant influence.
Then, $\phi_j$ is determined by maximizing the correlation coefficient between the sine wave in Eq. (4) and the reference waveform  for each voiced segment.
At the generation stage, $\phi_j$ is set as a random initial phase.
The DNN-transformed Gaussian white noise can be calculated offline for better run-time efficiency.

\begin{figure}[t]
    \centering
    \includegraphics[height=7.5cm]{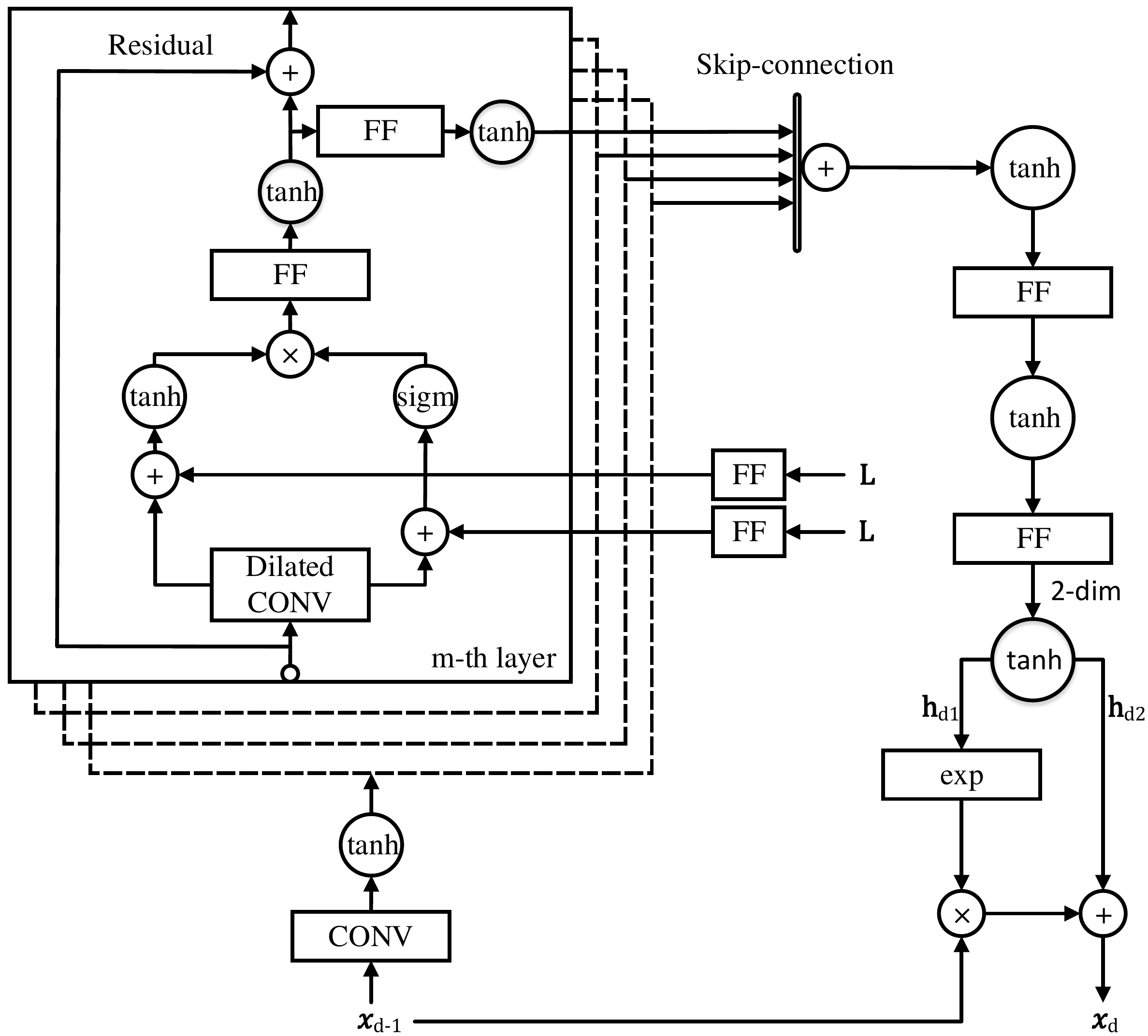}
    \caption{The structure of the $d$-th QWN block in the PSP. Here, \emph{FF}, \emph{CONV}, \emph{Dilated CONV} represent the feed-forward, convolutional and dilated convolutional layers respectively, \emph{tanh} and \emph{sigm} represent the hyperbolic tangent and sigmoid functions respectively.}
    \label{fig: QWN}
\end{figure}

\subsubsection{Filter Module}
\label{subsec2B2: Filter Module}

The excitation signal $\bm{e}$ generated by the source module and the upsampled LAS sequence $\bm{L}=[\bm{l}_1,\dots,\bm{l}_T]$ are input into the filter module.
Before upsampling, the frame-level LAS features pass through GRU-based recurrent layers and feed-forward layers for pre-processing. 
The output of the filter module is the predicted waveform $\hat{\bm{x}}$.

As shown in Fig. \ref{fig: PSP}, the filter module is a concatenation of $D$ quasi WaveNet (denoted by QWN) blocks.
Assume $\bm{x}_0=\bm{e}$ and $\bm{x}_D=\hat{\bm{x}}$.
The $d$-th QWN uses sequence $\bm{x}_{d-1}$ and $\bm{L}$ as input and predicts sequence $\bm{x}_{d}$.
The detailed structure of the $d$-th QWN is illustrated in Fig. \ref{fig: QWN}.
A QWN block is similar to a WaveNet model \cite{oord2016wavenet} whose key elements include dilated convolutions, gated activation units, residual connections and skip connections.
The difference is that QWNs are non-autoregressive with non-causal convolution because the whole sequence $\bm{x}_{d-1}$ is already known for the $d$-th block.
The LAS features are connected to the gated activation units after passing through two FF layers.
The hyperbolic tangent activation function is used in QWNs because the range of waveform samples is from -1 to 1.
The two FF layers after the skip connections are employed to reduce the dimensionality of the skip output and to generate sequences $\bm{h}_{d1}$ and $\bm{h}_{d2}$ both with length $T$.
Finally, the output sequence $\bm{x}_d$ is calculated as
\begin{align}
\label{equ: x_i}
\bm{x}_d=\bm{x}_{d-1}\odot\exp(\bm{h}_{d1})+\bm{h}_{d2},
\end{align}
where $\odot$ represents element-wise product.
The output of the last QWN $\bm{x}_D=\hat{\bm{x}}$ is used to define the loss function at the training stage and to extract phase spectra at the generation stage.

\subsubsection{Model Training}
\label{subsec2B3: Training Criteria}
Three loss functions are defined between the predicted waveform $\hat{\bm{x}}$ and the natural reference $\bm{x}$, including amplitude spectrum loss, waveform loss and negative correlation coefficient loss.
Comparing with the original NSF vocoder \cite{wang2019neural}, the last two losses are added for indirectly evaluating the phase accuracy of the predicted waveforms.

The amplitude spectrum loss is the MSE between the natural amplitude spectra and the predicted ones which are derived from $\bm{x}$ and $\hat{\bm{x}}$ using STFT respectively.
Similar to the NSF vocoder \cite{wang2019neural}, multiple sets of frame length ($FL$), frame shift ($FS$), and FFT point number ($FN$) are adopted for STFT in our implementation.
For the $i$-th set of ($FL^i$,$FS^i$,$FN^i$), the amplitude spectrum loss is calculated as
\begin{align}
\label{equ: L_ASi}
\mathcal L_{ASi}=\dfrac{1}{N^iK^i}\sum_{n=1}^{N^i}\sum_{k=1}^{K^i}(A_{n,k}^i-\hat{A}_{n,k}^i)^2,
\end{align}
where $A_{n,k}^i$ and $\hat{A}_{n,k}^i$ are the spectral amplitude at frame $n$ and frequency bin $k$ of $\bm{x}$ and $\hat{\bm{x}}$ respectively,
$N^i$ denotes the total number of frames and $K^i=\frac{FN^i}{2}+1$.

The waveform loss is defined as the MSE between the natural waveform samples and the predicted ones, i.e.,
\begin{align}
\label{equ: L_W}
\mathcal L_{W}=\dfrac{1}{T}\sum_{t=1}^{T}(x_t-\hat{x}_t)^2.
\end{align}

The negative correlation coefficient loss is calculated as the negative correlation coefficient between the natural waveform and the predicted waveform, i.e.,
\begin{align}
\label{equ: L_C}
\mathcal L_{C}=-\dfrac{\mathbb{E}[(\bm{x}-\mathbb{E}(\bm{x}))(\hat{\bm{x}}-\mathbb{E}(\hat{\bm{x}}))]}{\sqrt{\mathbb{V}(\bm{x})\mathbb{V}(\hat{\bm{x}})}},
\end{align}
where the functions $\mathbb{E}(\cdot)$ and $\mathbb{V}(\cdot)$ calculate mean and variance respectively.

Finally, the training criterion of the waveform generator in PSP is to minimize the combined loss function as
\begin{align}
\label{equ: L_PSP}
\mathcal L_{Comb}=\sum_{i}\mathcal L_{ASi}+\mathcal L_{W}+\mathcal L_{C}.
\end{align}

Similar with GAN-based ASP, we also try to use GANs to further improve the performance of PSP.
For GAN-based PSP, the generator (denoted by $G_P$) is the PSP model shown in Fig. \ref{fig: PSP} and the discriminator (denoted by $D_P$) is the same with that of GAN-based ASP shown in Fig. \ref{fig: ASP}.
The fake input of the discriminator is the output of the generator $\hat{\bm{x}}=G_P(\bm{f},\bm{L})$.
However, the real input is a reconstructed waveform $\bm{x}^*$ whose phase spectra are extracted from the natural waveform $\bm{x}$ but amplitude spectra are extracted from the predicted waveform generated by a well-trained PSP model.
There are two reasons.
One is that the PSP model always generates waveforms with poor amplitude spectra so the discriminator can easily distinguish the real and fake inputs if using natural waveform $\bm{x}$ as the real input.
Another is that the purpose of PSP is to generate excellent phase spectra so GAN only needs to further reduce the distance between the generated phase spectra and the natural ones.

A Wasserstein GAN \cite{gulrajani2017improved } loss with gradient penalty \cite{mescheder2018training} is also used for training the GAN-based PSP model.
The loss function of the discriminator is defined as
\begin{align}
\label{equ: PSP-D-loss}
\begin{split}
\mathcal L_{P-D}&=-\mathbb{E}_{\bm{x}^*\sim P^*_r}[D_P(\bm{x}^*)]+\mathbb{E}_{\hat{\bm{x}}\sim P_g}[D_P(\hat{\bm{x}})]\\
&+\lambda_{P-GP}\cdot\mathbb{E}_{\bm{x}^*\sim P^*_r,\hat{\bm{x}}\sim P_g}[(\max\{0,\Vert\nabla_{\tilde{\bm{x}}}D_P(\tilde{\bm{x}})\Vert-1\})^2],
\end{split}
\end{align}
where $\tilde{\bm{x}}=\varepsilon \bm{x}^*+(1-\varepsilon)\hat{\bm{x}}$ is sampled randomly along the line segment between $\bm{x}^*$ and $\hat{\bm{x}}$ and $\lambda_{P-GP}$ is a hyperparameter.
The loss function of the generator is defined as
\begin{align}
\label{equ: PSP-G-loss}
\mathcal L_{P-G}=-\mathbb{E}_{\hat{\bm{x}}\sim P_g}[D_P(\hat{\bm{x}})]+\lambda_{Comb}\cdot\mathcal L_{Comb},
\end{align}
where $\lambda_{Comb}$ is a hyperparameter.
Different from GAN-based ASP, the training process for GAN-based PSP is using $\mathcal L_{P-G}$ and $\mathcal L_{P-D}$ to train the generator and discriminator alternately like a standard GAN training process.

\section{Experiments}
\label{sec3: Experiments}

\subsection{Experimental Setup}
\label{sec3A: Experimental Setup}

The recordings of the female speaker \emph{slt} and the male speaker \emph{bdl} in CMU-ARCTIC databases \cite{kominek2004cmu} which contained English speech with 16 kHz sampling rate and 16bits resolution were adopted in our experiments.
For each speaker, we chose 1000 and 66 utterances to construct the training set and the validation set respectively, and the remaining 66 utterances were used as the test set.
The acoustic features at each frame were 43-dimensional including 40-dimensional mel-cepstra, an energy, an F0 and a voiced/unvoicded (V/UV) flag.
The natural acoustic features were extracted by STRAIGHT and the window size was 400 samples (i.e., 25ms) and the window shift was 80 samples (i.e., 5ms).
This paper focuses on neural vocoders, thus a simple acoustic model for SPSS was used in our experiments.
A bidirectional LSTM-RNN acoustic model \cite{fan2014tts} having 2 hidden layers with 1024 units per layer (512 forward units and 512 backward units) was trained to predict acoustic features from linguistic features.
The input linguistic features were the same as the ones used in Merlin toolkit \cite{wu2016merlin} for CMU-ARCTIC databases which were 425-dimensional.
The output of the acoustic model contained the acoustic features together with their delta and acceleration counterparts, which were totally 127 dimensions (the V/UV flag had no dynamic components).
Then, the predicted acoustic features were generated from the output by maximum likelihood parameter generation (MLPG) algorithm \cite{tokuda2000speech} considering global variance (GV) \cite{toda2007speech}.

Seven vocoders were compared in our experiments\footnote{Examples of generated speech can be found at \url{http://home.ustc.edu.cn/~ay8067/IEEETran_2019/demo.html}.}.
The descriptions of these vocoders are as follows.

\begin{itemize}
\item \emph{\textbf{STRAIGHT}} The conventional STRAIGHT vocoder.
At synthesis time, the spectral envelope at each frame was first reconstructed from input mel-cepstra and frame energy, and was then used to generate speech waveforms together with input source parameters (i.e., F0 and V/UV flag) \cite{kawahara1999restructuring}.

\item \emph{\textbf{WaveNet}} A 16-bit WaveNet-based neural vocoder, which is the teacher model used in parallel WaveNet \cite{oord2017parallel}.
Two speaker-dependent vocoders were trained using an open source implementation\footnote{
\url{https://github.com/r9y9/wavenet_vocoder}.}.
3 upsampling layers with upsampling rates \{5,4,4\}  were adopted.
Other configurations remained the same as that of the open source implementation.
The built model was a mixture density network, outputting the parameters for a mixture of 10 logistic distributions at each timestep and had 24 dilated casual convolutional layers which were divided into 4 convolutional blocks.
Each block contained 6 layers and their dilation coefficients were $\{2^0,2^1,\dots,2^5\}$.
The filter width was 3.
The number of gate channels in gated activation units was 512.
For the residual architectures, the number of residual channels was 512 and the number of skip channels was 256.
An \emph{Adam} optimizer \cite{kingma2014adam} was used to update the parameters by minimizing the negative log likelihood.
Models were trained and evaluated on a single Nvidia 1080Ti GPU using PyTorch framework \cite{paszke2017pytorch}.

\item \emph{\textbf{WaveRNN}} A 16-bit WaveRNN-based neural vocoder implemented by ourselves.
The structure was the same as the one used in our previous work \cite{ai2019dnn} which did not adopt the efficiency optimization strategies introduced in \cite{kalchbrenner2018efficient}.
The built model had one hidden layer of 1024 nodes where 512 nodes for coarse outputs and another 512 nodes for fine outputs.
The waveform samples were quantized to discrete values by 16-bit linear quantization.
Truncated back propagation through time (TBPTT) algorithm was employed to improve the efficiency of model training and the truncated length was set to 480.
An \emph{Adam} optimizer \cite{kingma2014adam} was used to update the parameters by minimizing the cross-entropy.
Models were trained and evaluated  on a single Nvidia 1080Ti GPU using TensorFlow framework \cite{abadi2016tensorflow}.

\item \emph{\textbf{NSF}} An NSF vocoder implemented by ourselves.
The model structure and the training method of the NSF vocoder were the same as that of the PSP model in the following HiNet vocoder \emph{\textbf{HiNet}}
except two main differences.
First, the NSF vocoder used acoustic features as input rather than LAS.
Second, log amplitude spectrum losses were added to its loss function and the loss function became $\mathcal L_{Comb}=\sum_{i}(\mathcal L_{LASi}+\lambda_{NSF}\cdot\mathcal L_{ASi})+\mathcal L_{W}+\mathcal L_{C}$, where $\lambda_{NSF}$ was set to 500 heuristically in our experiments.
To be consistent with the PSP model in HiNet vocoders, the strategy of using separate source-filter pairs for the harmonic and noise components of waveforms \cite{wang2019neural2} was not adopted here.

\item \emph{\textbf{HiNet}} Our proposed HiNet neural vocoder.
GANs were not used in both ASP and PSP.
When extracting LAS, the frame length and frame shift of STFT were 640 samples (i.e., 40ms) and 80 samples (i.e., 5ms) respectively and the FFT point number was 1024.
For ASP, the acoustic features at current frame along with 5 previous frames (i.e., $n_p=5$) were concatenated to form the complete input which was 258-dimensional.
There were two hidden layers with 2048 nodes per layer, and a 513-dimensional linear output layer which predicted the LAS at current frame.
The activation function was rectified linear units (ReLu) for hidden layers.
For PSP, an unidirectional GRU layer with 1024 nodes and an FF layer with 128 nodes were used to pre-process LAS.
When extracting phase spectra from the predicted waveform at the generation stage, the STFT parameter settings were consistent with the ones used for extracting LAS.
In the source module, the DNN for transforming Gaussian noise had two FF layers with 512 nodes per layer and hyperbolic tangent activation function together with a 1-dimensional linear output layer.
Hyperparameters $\alpha$ and $\sigma$ were set as 0.1 and 0.003 respectively.
Referring to the configuration of original NSF model \cite{wang2019neural}, the filter module consisted of 5 QWN blocks (i.e., $D=5$).
Each QWN had a non-causal convolutional layer for processing the input sequence and 10 dilated non-casual convolution layers and their dilation coefficients were $\{2^0,2^1,\dots,2^9\}$.
The filter width was 5.
The number of gate channels in gated activation units was 128.
The additional inputs $\bm{L}$ were connected to the gated activation units after passing through two FF layers both having 128 nodes.
For the residual architectures, the number of residual channels was 128 and the number of skip channels was 256.
After the skip connections, an FF layer with 16 nodes and an FF layer with 2 nodes were used to reduce the dimensionality of the skip output.
For the loss function of PSP, two sets of STFT configurations $(FL,FS,FN)$, i.e., $(320,80,512)$ and $(80,40,128)$,  were used for the amplitude spectrum loss. 
An \emph{Adam} optimizer \cite{kingma2014adam} was used to update the parameters by minimizing $\mathcal L_{MSE}$ and $\mathcal L_{Comb}$ for ASP and PSP respectively.
Truncated waveform sequences with 16000 samples were used for training PSP to avoid the overflow of GPU memory.
The initial learning rate of ASP was 0.0001 and the learning rate decreased exponentially from the 20th epoch.
The initial learning rate of PSP was also 0.0001 but the learning rate decreased exponentially from 2nd epoch.
Models were trained and evaluated on a single Nvidia 1080Ti GPU using TensorFlow framework \cite{abadi2016tensorflow}.

\item \emph{\textbf{HiNet-S}} A HiNet vocoder after PSP model simplification.
The simplified HiNet vocoder reduced the number of QWN blocks in PSP from 5 to 1, and halved the numbers of the gate channels, residual channels and skip channels compared with \emph{\textbf{HiNet}}.

\item \emph{\textbf{HiNet-S-GAN}} A simplified HiNet vocoder in which ASP and PSP both used GANs.
For GAN-based ASP, the generator was the ASP model in \emph{\textbf{HiNet-S}}.
The discriminator consisted of 6 convolutional layers (filter width=7, stride size=2) as shown in Fig. \ref{fig: ASP} and their channels were 16, 32, 64, 128, 256 and 512 respectively.
The resulting dimensions per layer, being it frequency bins $\times$ channels, were 513$\times$1, 257$\times$16, 129$\times$32, 65$\times$64, 33$\times$128, 17$\times$256 and 9$\times$512.
Finally, two FF layers with 512 and 9 nodes respectively were used to map the 9$\times$512 convolutional results into a value for loss calculation.
There were three steps in the training process.
(1) The generator was first trained for 25 epochs. The initial learning rate was 0.0002 and the learning rate decreased exponentially from the 20th epoch.
(2) The discriminator was then trained for 15 epochs. The initial learning rate was 0.00005 and the learning rate decreased exponentially from the 10th epoch.
(3) The generator and the discriminator were alternately trained. The initial learning rates for the generator and the discriminator were 0.0002 and 0.00005 respectively and the learning rates both decreased exponentially from the 2nd epoch.
Hyperparameters $\lambda_{A-GP}$ and $\lambda_{MSE}$ were set as 10 and 50 respectively.
For GAN-based PSP, the generator was the PSP model in \emph{\textbf{HiNet-S}}.
The discriminator consisted of 11 convolutional layers (filter width=31, stride size=2) as shown in Fig. \ref{fig: ASP} and their channels were 16, 32, 32, 64, 64, 128, 128, 256, 256, 512 and 1024 respectively.
The resulting dimensions per layer, being it samples$\times$channels, were 16000$\times$1, 8000$\times$16, 4000$\times$32, 2000$\times$32, 1000$\times$64, 500$\times$64, 250$\times$128, 125$\times$128, 63$\times$256, 32$\times$256, 16$\times$512 and 8$\times$1024.
Finally, two FF layers with 1024 and 8 nodes respectively were used to map the 8$\times$1024 convolutional results into a value for loss calculation.
The initial learning rates of the generator and the discriminator were both 0.0001 and the learning rates both decreased exponentially from
the 2nd epoch.
Hyperparameters $\lambda_{P-GP}$ and $\lambda_{Comb}$ were both 10.

\end{itemize}

\subsection{Comparison between HiNet and Some Existing Vocoders}
\label{sec3B: Comparison among HiNet Vocoder and Existing Vocoders}

In this section, we compared the performance of our proposed \emph{\textbf{HiNet}} vocoder with three representative existing vocoders, including \emph{\textbf{STRAIGHT}}, \emph{\textbf{WaveNet}} and \emph{\textbf{WaveRNN}} 
by objective and subjective evaluations.

First, we compared the distortions between natural speech and the speech reproduced by these four vocoders when using natural acoustic features as input.
Five objective metrics used in \cite{tamamori2017speaker} were adopted here, including signal-to-noise ratio (SNR) which reflected the distortion
of waveforms, root MSE (RMSE) of LAS (denoted by LAS-RMSE) which reflected the distortion in frequency domain, mel-cepstrum distortion (MCD) which described the distortion of mel-cepstra, MSE of F0 which reflected the distortion of F0 (denoted by F0-RMSE), and V/UV error which was the ratio between the number of frames
with mismatched V/UV flags and the total number of frames.
Among these metrics, SNR can be considered as an overall measurement on the distortions of both amplitude and phase spectra, while LAS-RMSE and MCD mainly present the distortion of amplitude spectra.
Besides, the SNR for voiced frames (denoted by SNR-V) was also calculated for each vocoder.
SNR-V can better present the distortion in phase spectra than the overall SNR because the unvoiced frames with random phase spectra were excluded when calculating SNR-V. 
STRAIGHT was used to extract acoustic features from both original and reproduced speech waveforms for calculating all these metrics.

\begin{table}
\centering
    \caption{Objective evaluation results of \emph{\textbf{STRAIGHT}}, \emph{\textbf{WaveNet}}, \emph{\textbf{WaveRNN}} and \emph{\textbf{HiNet}} on the test sets of two speakers when using natural acoustic features as input.}
    \resizebox{8.8cm}{2.2cm}{
    \begin{tabular}{c c | c c c c}
        \hline
        \hline
         & & \emph{\textbf{STRAIGHT}}& \emph{\textbf{WaveNet}}& \emph{\textbf{WaveRNN}} & \emph{\textbf{HiNet}}\\
         \hline
         \multirow{6}{*}{\emph{slt}}
         & SNR(dB) & 0.5357 & 3.5228 & 6.0568 & \textbf{6.2937}\\
         & SNR-V(dB) & 1.3551 & 5.3285 & 8.6591 &  \textbf{8.9254}\\
         & LAS-RMSE(dB) & \textbf{5.5800} & 6.0681 & 6.2489 & 5.5937\\
         & MCD(dB) & \textbf{1.3315} & 1.5950 & 1.6042 & 1.5036\\
         & F0-RMSE(cent) & 14.8430 & 71.9886 & 12.1309 & \textbf{8.0286}\\
         & V/UV error(\%) & 3.3994 & 4.6260 & 3.3756 & \textbf{2.1971}\\
         \cline{1-6}
         \multirow{6}{*}{\emph{bdl}}
         & SNR(dB) & 1.0987 & 2.7105 & 3.8993 & \textbf{4.5905}\\
         & SNR-V(dB) & 2.2865 & 3.9987 & 5.8108 &  \textbf{6.5571}\\
         & LAS-RMSE(dB) & \textbf{5.6434} & 6.0581 & 6.1812 & 5.7486\\
         & MCD(dB) & \textbf{1.3097} & 1.4093 & 1.5150 & 1.5528\\
         & F0-RMSE(cent) & 25.7898 & 98.3218 & 21.0020 & \textbf{10.5880}\\
         & V/UV error(\%) & 4.5588 & 8.7091 & 5.5817 & \textbf{2.7663}\\
        \hline
        \hline
    \end{tabular}}
\label{tab_four_vocoders}
\end{table}

\begin{figure*}[t]
    \centering
    \includegraphics[height=5.5cm]{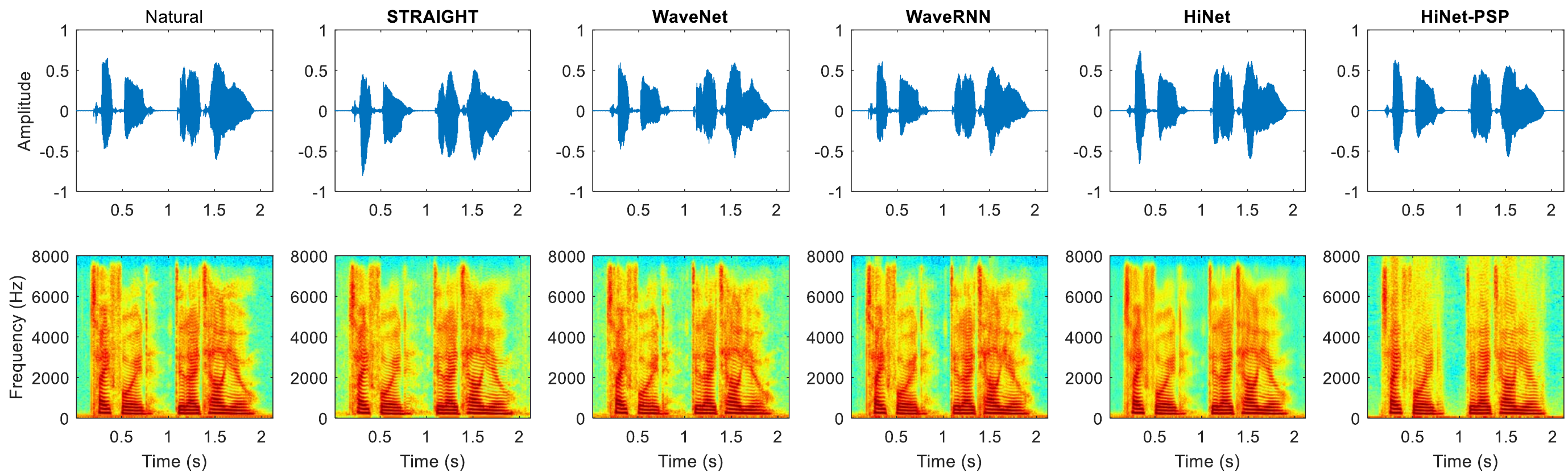}
    \caption{The waveforms and spectrograms of natural speech and the speech generated by different vocoders when using natural acoustic features as input for an example sentence (arctic\_b0536) in the test set of speaker \emph{slt}. Here,  \emph{HiNet-PSP} denotes the waveforms generated by the PSP in HiNet. }
    \label{fig: Waveform and Spectrogram}
\end{figure*}

The results on the test sets of the two speakers are listed in Table \ref{tab_four_vocoders}.
It is obvious that \emph{\textbf{STRAIGHT}} achieved the lowest SNR and SNR-V for both speakers due to the neglect of natural phase information.
Our proposed \emph{\textbf{HiNet}} vocoder outperformed \emph{\textbf{WaveNet}} and \emph{\textbf{WaveRNN}} on the SNR and SNR-V metric for both speakers.
This indicated that the HiNet vocoder restored the shape of waveforms more accurately than other vocoders.
Besides, our proposed \emph{\textbf{HiNet}} vocoder achieved the lowest LAS-RMSE among the three neural vocoders
which implied the advantage of using a separate ASP in our proposed method.
Regarding with MCD, the results on these two speakers were inconsistent which needed further investigation.
Our proposed \emph{\textbf{HiNet}} vocoder achieved the lowest F0-RMSE and V/UV error among all four vocoders and their differences were significant.
This advantage can be attributed to the explicit excitation signal determined by F0s and U/V flags in the PSP of \emph{\textbf{HiNet}}.

Fig. \ref{fig: Waveform and Spectrogram} shows the waveforms and spectrograms of natural speech and the speech generated by different vocoders when using natural acoustic features as input for an example sentence in the test set of speaker \emph{slt}.
We can see that there was observable difference between the overall contours of the waveforms generated by \emph{\textbf{STRAIGHT}} and the natural waveforms due to the neglect of natural phase information in \emph{\textbf{STRAIGHT}}.
In contrast, the neural vocoders (i.e., \emph{\textbf{WaveNet}}, \emph{\textbf{WaveRNN}} and \emph{\textbf{HiNet}}) restored the overall waveform contours much better.
Besides, our proposed \emph{\textbf{HiNet}} vocoder was better at reconstructing the high-frequency harmonic structures of some voiced segments (e.g., 1.4$\sim$1.6s and 4000$\sim$6000Hz in Fig. \ref{fig: Waveform and Spectrogram})
as shown in the spectrograms.

\begin{table}
\centering
    \caption{Real time factors (RTFs) on GPU and CPU and number of model parameters (NMP) of compared vocoders, where ($\cdot$/$\cdot$) represents the frame-level NMP (left) and point-level NMP (right).}
    \resizebox{9.0cm}{0.8cm}{
    \begin{tabular}{c | c c c c}
        \hline
        \hline
         Vocoder& \emph{\textbf{WaveNet}}& \emph{\textbf{WaveRNN}}& \emph{\textbf{HiNet}} & \emph{\textbf{HiNet-S}} \\
         \hline
         RTF (GPU) & 222.3656 & 100.9148 & 0.3420 & \textbf{0.1929}\\
         RTF (CPU) & 510.6052 & 900.4469 & 38.2128 & \textbf{11.5570}\\
         NMP ($\times10^6$) & 72.9 (0/72.9)& \textbf{5.3} (0/5.3) & 23.0 (10.6/12.4) &11.2 (10.6/\textbf{0.6})\\
        \hline
        \hline
    \end{tabular}}
\label{tab_RTF}
\end{table}

\begin{figure}[t]
    \centering
    \includegraphics[height=4cm]{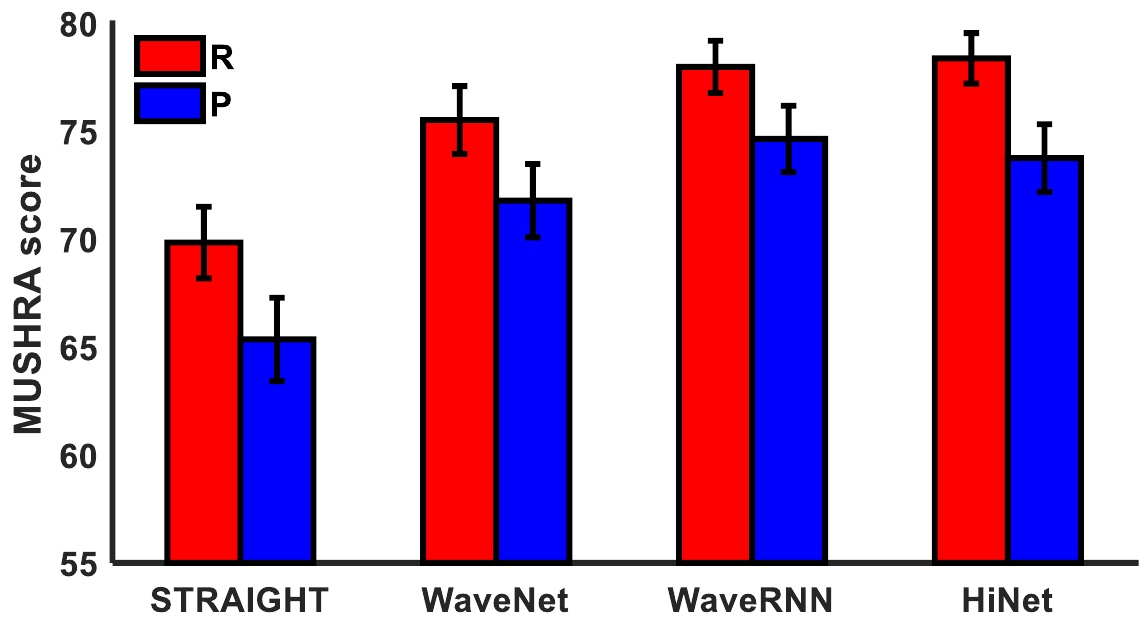}
    \caption{Average MUSHRA scores with 95\% confidence interval of the four vocoders for speaker \emph{slt}. ``R" stands for using natural acoustic features as input and ``P" stands for using predicted acoustic features as input.}
    \label{fig: MUSHRA_slt}
\end{figure}

\begin{figure}[t]
    \centering
    \includegraphics[height=4cm]{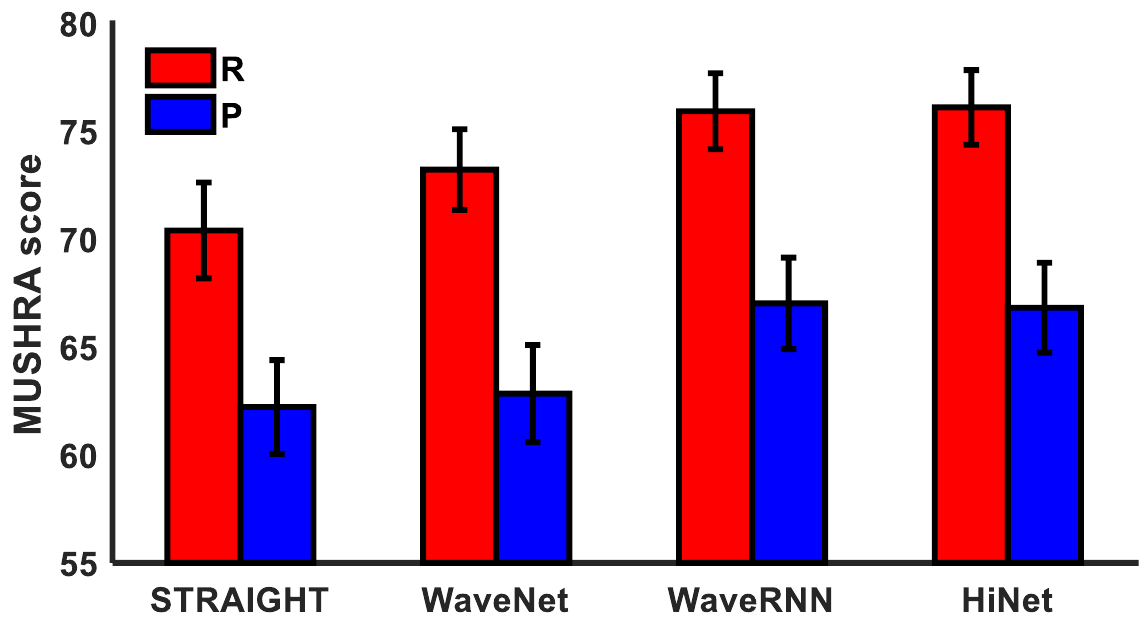}
    \caption{Average MUSHRA scores with 95\% confidence interval of the four vocoders for speaker \emph{bdl}. ``R" stands for using natural acoustic features as input and ``P" stands for using predicted acoustic features as input.}
    \label{fig: MUSHRA_bdl}
\end{figure}

In order to evaluate the run-time efficiency of different neural vocoders, real time factor (RTF) which is defined as the ratio between the time consumed to generate speech waveforms and the duration of the generated speech was utilized as the measurement.
In our implementation, the RTF value was calculated as the ratio between the time consumed to generate all test sentences using a single Nvidia 1080Ti GPU or a single CPU core and the total duration of the test set.
The results are listed in Table \ref{tab_RTF}.
It can be observed that our proposed \emph{\textbf{HiNet}} vocoder achieved the highest generation efficiency, no matter using GPU or CPU.
For \emph{\textbf{WaveNet}} and \emph{\textbf{WaveRNN}}, they were very inefficient due to the point-by-point autoregressive generation.

The last row of Table \ref{tab_RTF} shows the number of model parameters (NMP).
We also calculated the number of model parameters operating at frame-level (frame-level NMP) and point-level (point-level NMP) respectively in Table \ref{tab_RTF}
because the run-time efficiency was much more sensitive to the point-level NMP than to the frame-level NMP.
For \emph{\textbf{WaveNet}}, only the layers for upsampling the input acoustic features were operated at frame-level but their number of parameters was so small that it can be ignored in the total NMP.
For \emph{\textbf{WaveRNN}}, the upsampling was operated by repeating and had no parameters.
Therefore, there were no frame-level NMPs for both \emph{\textbf{WaveNet}} and \emph{\textbf{WaveRNN}} as shown in Table \ref{tab_RTF}.
For \emph{\textbf{HiNet}}, all point-level model parameters  existed in the filter module of the PSP.
Other modules of the PSP and the whole ASP operated at frame-level.
It can be observed that both the total NMP and point-level NMP of \emph{\textbf{HiNet}} were much smaller than that of \emph{\textbf{WaveNet}} but still larger than that of \emph{\textbf{WaveRNN}}.
A further analysis showed that 94.4\% and 98.4\% of the time used by \emph{\textbf{HiNet}} was spent on the PSP when using GPU and CPU respectively.
This inspires us to simplify the waveform generator in PSP  for further improving the efficiency and reducing the NMP of \emph{\textbf{HiNet}}.

Regard with the subjective evaluation, four MUSHRA (MUltiple Stimuli with Hidden Reference and Anchor) tests \cite{recommendation2001method} were conducted to compare the naturalness of these four vocoders with natural recordings as references for both speakers and using both natural and predicted acoustic features as input.
In each test, twenty test sentences synthesized by the four vocoders were evaluated by at least 30 English native listeners on the crowdsourcing platform of Amazon Mechanical
Turk\footnote{\url{https://www.mturk.com}.} with anti-cheating considerations \cite{buchholz2011crowdsourcing}.
Listeners were asked to give a naturalness score between 0 and 100 to each sample and the reference natural recording had the maximum score of 100.

The average naturalness scores and their 95\% confidence intervals of these four vocoders are shown in Fig. \ref{fig: MUSHRA_slt} and Fig. \ref{fig: MUSHRA_bdl} for speaker \emph{slt} and \emph{bdl} respectively.
The results of paired $t$-test showed that our proposed \emph{\textbf{HiNet}} vocoder outperformed \emph{\textbf{STRAIGHT}} and \emph{\textbf{WaveNet}} significantly at significance level of 0.01 and the differences between \emph{\textbf{HiNet}} and \emph{\textbf{WaveRNN}} were not significant for both speakers, no matter using natural or predicted acoustic features as input.
Besides, the differences between \emph{\textbf{STRAIGHT}} and \emph{\textbf{WaveNet}} for speaker \emph{bdl} when using predicted acoustic features as input were also not significant.
This may be attributed to the severe F0 distortion  (F0-RMSE=122.6858 cent) of \emph{\textbf{WaveNet}} for speaker \emph{bdl} when using predicted acoustic features as input.
Although our proposed \emph{\textbf{HiNet}} vocoder achieved  similar performance with that of \emph{\textbf{WaveRNN}},
 its run-time efficiency was about 300 times higher as shown in Table \ref{tab_RTF}.

\subsection{Comparison between HiNet and NSF Vocoders}
\label{sec3B: Comparison between HiNet and NSF Vocoders}

In this section, we compared the performance of four vocoders including \emph{\textbf{HiNet}}, \emph{\textbf{NSF}}, \emph{\textbf{HiNet-S}} and \emph{\textbf{HiNet-S-GAN}} mentioned in Section \ref{sec3A: Experimental Setup} by objective and subjective evaluations.
These four vocoders were based on similar models (i.e., \emph{\textbf{NSF}}) or variants of \emph{\textbf{HiNet}} (i.e., \emph{\textbf{HiNet-S}} and \emph{\textbf{HiNet-S-GAN}}).

\begin{table}
\centering
    \caption{Objective evaluation results of \emph{\textbf{HiNet}}, \emph{\textbf{NSF}} and \emph{\textbf{HiNet-S}} on the test sets of two speakers when using natural acoustic features as input.}
    \begin{tabular}{c c | c c c}
        \hline
        \hline
         & & \emph{\textbf{HiNet}}& \emph{\textbf{NSF}}& \emph{\textbf{HiNet-S}}\\
         \hline
         \multirow{7}{*}{\emph{slt}}
         & SNR(dB) & 6.2937 & \textbf{6.8664} & 6.1952 \\
         & SNR-V(dB) & 8.9254 & \textbf{10.1387} & 8.8350 \\
         & LAS-RMSE(dB) & \textbf{5.5937} & 6.5548 & 5.6291 \\
         & MCD(dB) & 1.5036 & 2.0470 & \textbf{1.5030} \\
         & F0-RMSE(cent) & 8.0286 & \textbf{7.0935} & 7.4500 \\
         & V/UV error(\%) & 2.1971 & 2.1319  & \textbf{2.1019} \\
         \cline{1-5}
         \multirow{7}{*}{\emph{bdl}}
         & SNR(dB) & 4.4062 & \textbf{4.4974} & 3.8400 \\
         & SNR-V(dB) & \textbf{6.5571} & 6.5201 & 5.8538 \\
         & LAS-RMSE(dB) & \textbf{5.9188} & 7.4945 & 6.0612 \\
         & MCD(dB) & \textbf{1.5081} & 2.4338 & 1.5678 \\
         & F0-RMSE(cent) & 11.8820 & \textbf{11.8750} & 12.3931 \\
         & V/UV error(\%) & \textbf{2.6708} & 3.2978 & 3.1686 \\
        \hline
        \hline
    \end{tabular}
\label{tab_three_vocoders}
\end{table}

\begin{table*}
\centering
    \caption{Average preference scores (\%) on speech quality among different vocoders of two speakers when using natural acoustic features as input, where N/P stands for ``no preference" and $p$ denotes the $p$-value of a $t$-test between two vocoders.}
    \begin{tabular}{c c|c c c c c c}
        \hline
        \hline
        & & \emph{\textbf{HiNet}}& \emph{\textbf{NSF}}& \emph{\textbf{HiNet-S}}& \emph{\textbf{HiNet-S-GAN}}
         &\emph{N/P}& $p$ \\
         \hline
         \multirow{3}{*}{\emph{slt}}
         &\emph{\textbf{HiNet}} vs \emph{\textbf{NSF}} & \textbf{50.47} & 25.31 & -- & --  & 24.22 & $<$0.01 \\
         &\emph{\textbf{HiNet}} vs \emph{\textbf{HiNet-S}} & 29.67 & -- & 30.66 & --  & 39.67 & 0.7528 \\
         &\emph{\textbf{HiNet-S}} vs \emph{\textbf{HiNet-S-GAN}} & -- & -- & 30.33 & \textbf{37.83} & 31.83 & 0.0260 \\
         \hline
         \multirow{3}{*}{\emph{bdl}}
         &\emph{\textbf{HiNet}} vs \emph{\textbf{NSF}} & \textbf{54.69} & 24.31 & -- & -- & 21.00 & $<$0.01 \\
         &\emph{\textbf{HiNet}} vs \emph{\textbf{HiNet-S}} & 31.13 & -- & 26.45 & -- & 42.42 & 0.1249 \\
         &\emph{\textbf{HiNet-S}} vs \emph{\textbf{HiNet-S-GAN}} & -- & -- & 28.23 & 29.03 & 42.74 & 0.7910 \\
         \hline
        \hline
    \end{tabular}
\label{tab: ABX}
\end{table*}

\begin{figure}
    \centering
    \includegraphics[height=3.1cm]{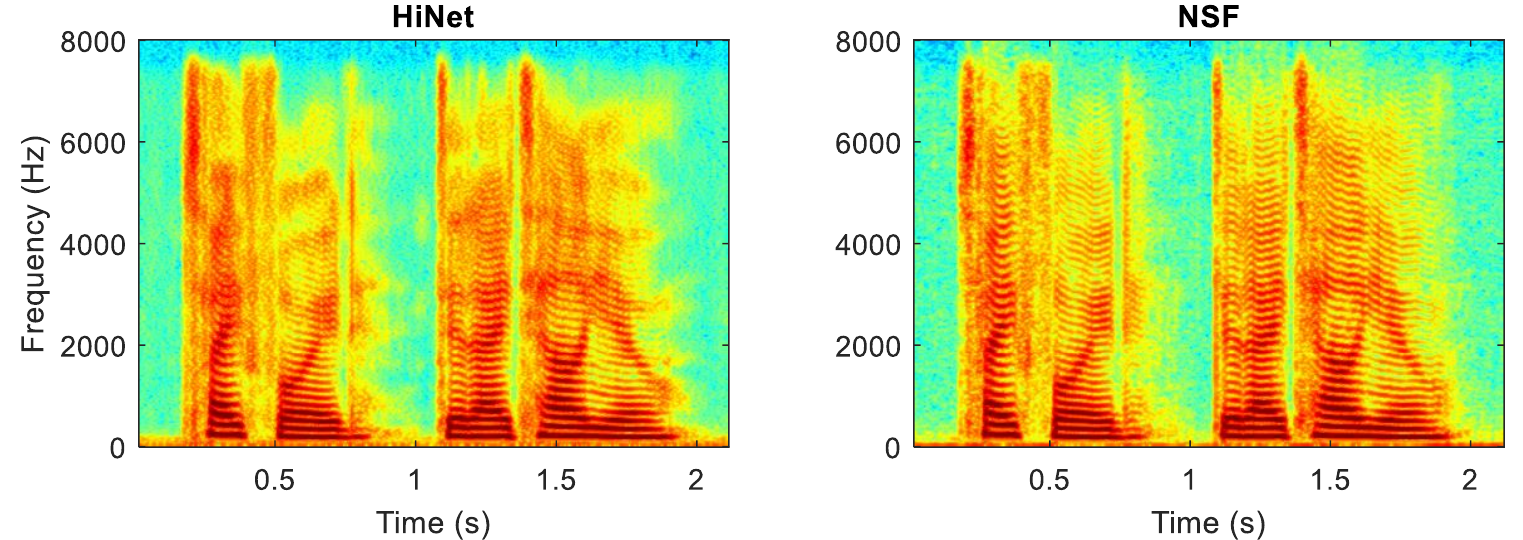}
    \caption{The spectrograms of the speech generated by \emph{\textbf{HiNet}} and \emph{\textbf{NSF}}
    when using natural acoustic features as input for an example sentence (arctic\_b0536) in the test set of speaker \emph{slt}.}
    \label{fig: Spectrogram_NSF}
\end{figure}

We first compared the performance of our proposed HiNet vocoder (i.e., \emph{\textbf{HiNet}}) and NSF vocoder (i.e., \emph{\textbf{NSF}}).
Our PSP model in the HiNet vocoder was inspired by the NSF vocoder.
When implementing the NSF vocoder, we found that only using log amplitude spectrum loss referring to the original paper \cite{wang2019neural} always caused very strong high-frequency harmonics, however only using amplitude spectrum loss always caused strong high-frequency noise.
Therefore, we adopt a combination of these two losses as shown in Section \ref{sec3A: Experimental Setup}.
Besides, we also found that the waveform loss and correlation coefficient loss and the pre-calculated initial phase were helpful to improve the performance of the NSF vocoder.
However, the waveforms generated by \emph{\textbf{NSF}} still had strong high-frequency harmonics as shown in the spectrograms of Fig. \ref{fig: Spectrogram_NSF}.

The objective results are listed in Table \ref{tab_three_vocoders}.
It is obvious that \emph{\textbf{NSF}} achieved better performance on SNR, SNR-V, F0-RMSE and V/UV error metrics but achieved worse performance on LAS-RMSE and MCD metrics than \emph{\textbf{HiNet}}.
This indicated that the phase spectra generated by the NSF vocoder were more precise but the amplitude spectra was the opposite.
We have also tried to use the NSF vocoder as PSP in the HiNet vocoder, but the results were unsatisfactory.

Regard with the subjective evaluation, two groups of ABX preference tests were conducted to examine whether there were significant subjective differences between the waveforms generated by \emph{\textbf{HiNet}} and \emph{\textbf{NSF}} when using natural acoustic features as input for both speakers.
In each subjective test, twenty sentences randomly selected from the test set were synthesised by two comparative vocoders.
Each pair of generated speech were evaluated by at least 30 English native listeners on the crowdsourcing platform of Amazon Mechanical Turk
in random order.
The listeners were asked to judge which utterance in each pair had better speech quality or there was no preference. In addition to calculating the average preference scores, the $p$-value of a $t$-test was used to measure the significance of the difference between two vocoders.
The results are listed in Table \ref{tab: ABX} (\emph{\textbf{HiNet}} vs \emph{\textbf{NSF}}).
It is obvious that \emph{\textbf{HiNet}} outperformed \emph{\textbf{NSF}} very significantly ($p < $0.01) for both speakers.
This may be attributed to the poor listening feeling caused by excessive high-frequency harmonics of the waveforms generated by \emph{\textbf{NSF}}.

Then, we explored whether a scale-reduced neural waveform generator was enough for predicting phase spectra in order to further decrease the computation complexity of the HiNet vocoder because the PSP model consumed most of the computation at the generation stage of the HiNet vocoder as mentioned in Section \ref{sec3B: Comparison among HiNet Vocoder and Existing Vocoders}.
The objective results are listed in Table \ref{tab_three_vocoders}.
By comparing \emph{\textbf{HiNet}} and \emph{\textbf{HiNet-S}}, we found that there were no significant degradations on all metrics after simplifying the structure of the neural waveform generator in PSP.
ABX preference tests also confirmed that there was
no significant difference ($p >$ 0.05) between the subjective quality of \emph{\textbf{HiNet}} and \emph{\textbf{HiNet-S}} for both speakers as shown in Table \ref{tab: ABX}.  
These results indicated that the performance of the HiNet vocoder was insensitive to the complexity of the NSF-based waveform generator in PSP to some extend. A neural waveform generator with much smaller scale than the ones for direct waveform generation may be enough for phase recovery.
Besides, we also compared the RTF and NMP between \emph{\textbf{HiNet}} and \emph{\textbf{HiNet-S}} as shown in Table \ref{tab_RTF}.
By simplifying the PSP model, the RTF decreased from 0.34 to 0.19 and from 38.21 to 11.56 when using GPU and CPU respectively and the generation efficiency was greatly improved.
Besides, the total NMP and point-level NMP of \emph{\textbf{HiNet-S}} were also reduced to $11.2\times10^6$ and $0.6\times10^6$ respectively.
The point-level NMP of \emph{\textbf{HiNet-S}} was only about one ninth of that of \emph{\textbf{WaveRNN}}.

\begin{figure}
    \centering
    \includegraphics[height=3.1cm]{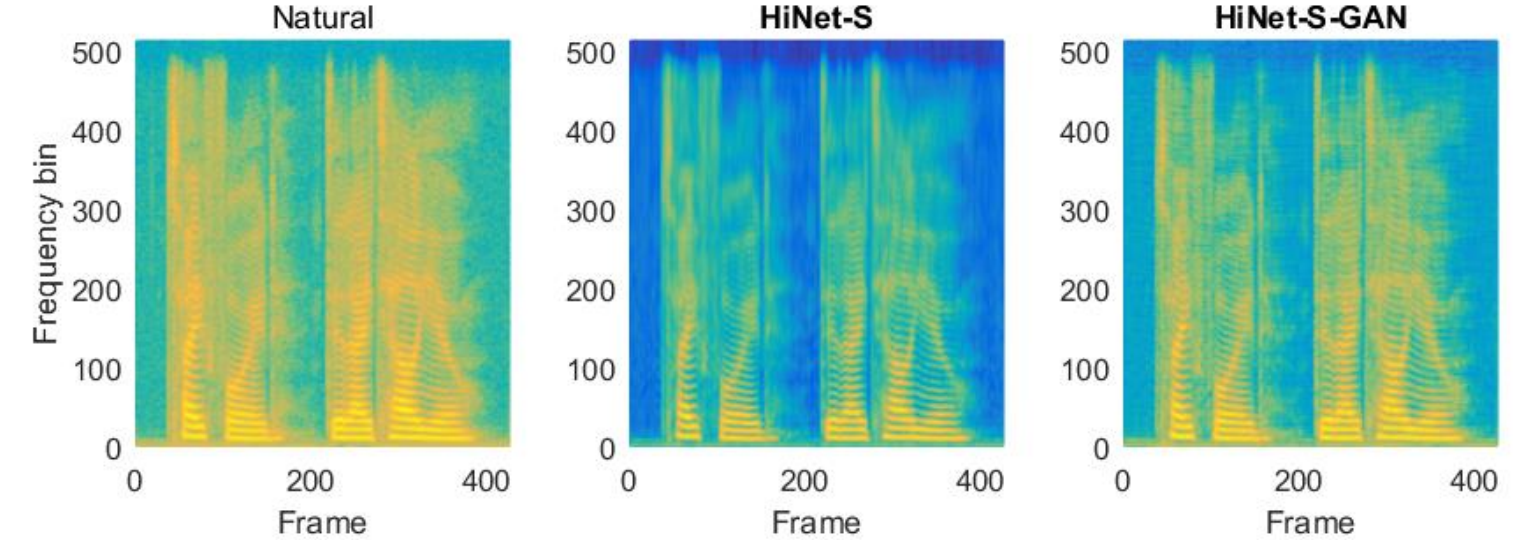}
    \caption{Color maps of the natural LAS and LAS generated by the ASP in \emph{\textbf{HiNet-S}} and \emph{\textbf{HiNet-S-GAN}}
    when using natural acoustic features as input for an example sentence (arctic\_b0536) in the test set of speaker \emph{slt}.}
    \label{fig: LAS_GAN}
\end{figure}

Finally, we explored whether GANs contribute to improving the performance of the HiNet vocoder by comparing \emph{\textbf{HiNet-S}} and \emph{\textbf{HiNet-S-GAN}}.
Here, the simplified HiNet vocoder was utilized.
Once GANs were used in the model, objective evaluations lost their effects.
Therefore, only subjective evaluations were conducted and the results of ABX tests are listed in Table \ref{tab: ABX} (\emph{\textbf{HiNet-S}} vs \emph{\textbf{HiNet-S-GAN}}).
For speaker \emph{slt}, \emph{\textbf{HiNet-S-GAN}} outperformed \emph{\textbf{HiNet-S}} significantly ($p <$ 0.05).
However, there was no significant difference ($p >$ 0.05) between these two vocoders for speaker \emph{bdl}.
Besides, we also draw the color maps of the natural LAS and LAS generated by the ASP in \emph{\textbf{HiNet-S}} and \emph{\textbf{HiNet-S-GAN}} as shown in Fig. \ref{fig: LAS_GAN}.
It is obvious that the MSE loss-based ASP in \emph{\textbf{HiNet-S}} suffered from the over-smoothing problem and lost some spectral details such as high-frequency formants.
In contrast, the GAN-based ASP in \emph{\textbf{HiNet-S-GAN}} alleviated the over-smoothing issue and the generated LAS were more similar with the natural ones.

\subsection{Discussions}
\label{sec3D: Discussions}

\subsubsection{Impact of the amount of training data on the HiNet vocoder}
\label{subsec3D1: Impact of the amount of training data on the HiNet vocoder}

\begin{figure}
    \centering
    \includegraphics[height=6.4cm]{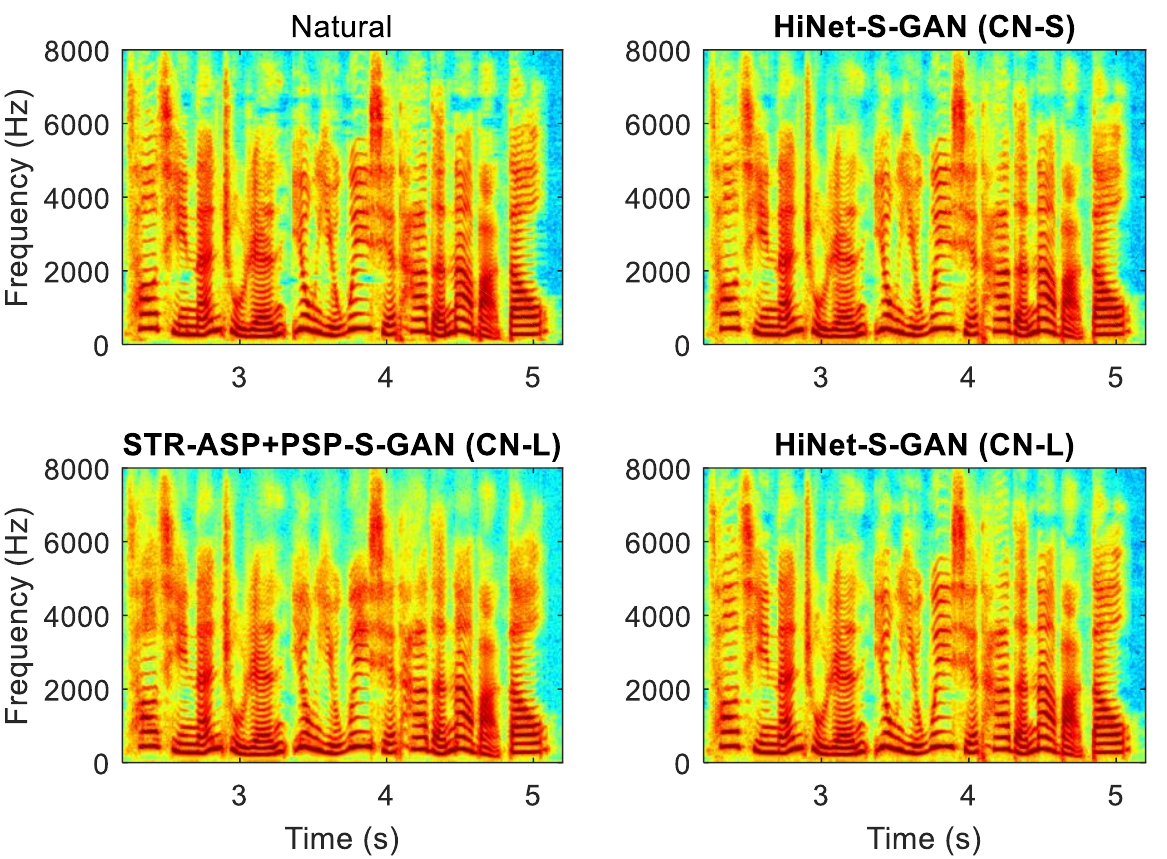}
    \caption{The spectrograms of natural speech, of the speech generated by the \emph{\textbf{HiNet-S-GAN}} vocoders built on the \emph{CN-S} and \emph{CN-L} datasets, and of the speech generated by the \emph{\textbf{STR-ASP+PSP-S-GAN}} vocoder built on the \emph{CN-L} dataset
    for an example sentence in the test set. Here, natural acoustic features were used  as input.}
    \label{fig: Spectrogram_STRAIGHT}
\end{figure}

In this subsection, we discussed the impact of the amount of training data on the performance of the HiNet vocoder.
It is worth mentioning that the speech data of single speaker in the Arctic dataset was only about 1 hour.
Therefore, a much larger dataset from a Chinese female speaker was used in this experiment. 
Two training sets were designed for building \emph{\textbf{HiNet-S-GAN}} vocoders. One contained 800 training utterances whose duration was 1.9 hours (denoted as \emph{CN-S}) and another contained 13134 training utterances whose duration was 19.5 hours (denoted as \emph{CN-L}).
The validation set and the test set had 100 utterances respectively.
An ABX subjective preference test was conducted between the two \emph{\textbf{HiNet-S-GAN}} vocoders trained on \emph{CN-S} and \emph{CN-L}.
10 Chinese native speakers participated in the test.
The results are listed in Table \ref{tab: ABX_CN}.
It is obvious that \emph{CN-L} outperformed \emph{CN-S} very significantly ($p <$ 0.01),
which indicates that the HiNet vocoder can benefit from a large training set.
We also drew the spectrograms generated by these two vocoders as shown in Fig. \ref{fig: Spectrogram_STRAIGHT}.
We can see that the spectrogram generated by \emph{\textbf{HiNet-S-GAN}}(\emph{CN-L}) was very close to the natural one.
Compared with \emph{\textbf{HiNet-S-GAN}}(\emph{CN-L}), \emph{\textbf{HiNet-S-GAN}}(\emph{CN-S}) lost some spectral details (e.g., 4.8$\sim$5.2s).

\begin{table}
\centering
    \caption{Average preference scores (\%) on speech quality between the \emph{\textbf{HiNet-S-GAN}} vocoders built on \emph{CN-S} and \emph{CN-L} datasets when using natural acoustic features as input, where N/P stands for ``no preference" and $p$ denotes the $p$-value of a $t$-test between two vocoders.}
    \begin{tabular}{c c c c}
        \hline
        \hline
        \emph{CN-S}& \emph{CN-L} & \emph{N/P}& $p$ \\
         \hline
         \textbf{58.00} & 7.00 & 35.00 & $<$ 0.01\\
         \hline
        \hline
    \end{tabular}
\label{tab: ABX_CN}
\end{table}

\subsubsection{Comparison between GAN-based ASP and conventional one}
\label{subsec3D2: Comparison between GAN-based ASP and conventional one}
As we all know, some conventional signal processing algorithms can also achieve the function of ASP, i.e., converting acoustic features into amplitude spectra.
For example, the STRAIGHT vocoder can be used as an ASP by extracting amplitude spectra from its generated waveforms. 
We designed a new vocoder, named \emph{\textbf{STR-ASP+PSP-S-GAN}}, which first extracted amplitude spectra from the waveforms generated by \emph{\textbf{STRAIGHT}} and then input the amplitude spectra into the GAN-based simplified PSP model used in \emph{\textbf{HiNet-S-GAN}}.
We compared the performance of \emph{\textbf{HiNet-S-GAN}} and \emph{\textbf{STR-ASP+PSP-S-GAN}} because these two vocoders just had different ASP.
The subjective ABX preference test results for speaker \emph{slt} are listed in Table \ref{tab: ABX_STRAIGHT}.
We can see that there was no significant difference ($p >$ 0.05) between these two vocoders for speaker \emph{slt}.

Considering the small training set (0.8 hour) of speaker \emph{slt}, we also conducted the comparative experiments on the \emph{CN-S} and \emph{CN-L} datasets
and the results are shown in Table \ref{tab: ABX_STRAIGHT}.
It is obvious that for both \emph{CN-S} and \emph{CN-L}, \emph{\textbf{HiNet-S-GAN}} outperformed \emph{\textbf{STR-ASP+PSP-S-GAN}} very significantly ($p <$ 0.01).
According to the listeners' feedback, it was easier to distinguish the differences when conducting the ABX test on the \emph{CN-L} dataset.
Fig. \ref{fig: Spectrogram_STRAIGHT} also shows that the spectrogram of \emph{\textbf{HiNet-S-GAN}} was more similar to the natural one than that of \emph{\textbf{STR-ASP+PSP-S-GAN}} when using the same \emph{CN-L} training set.
These results indicated that the GAN-based ASP can outperform the STRAIGHT-based one given enough training data.

\begin{table}
\centering
    \caption{Average preference scores (\%) on speech quality between \emph{\textbf{HiNet-S-GAN}} and \emph{\textbf{STR-ASP+PSP-S-GAN}} for \emph{slt} (0.8 hour), \emph{CN-S} (1.9 hours) and \emph{CN-L} (19.5 hours) datasets when using natural acoustic features as input, where N/P stands for ``no preference" and $p$ denotes the $p$-value of a $t$-test between two vocoders.}
    \begin{tabular}{c  c c c c}
        \hline
        \hline
        & \emph{\textbf{HiNet-S-GAN}}& \emph{\textbf{STR-ASP+PSP-S-GAN}} & \emph{N/P}& $p$ \\
         \hline
         \emph{slt} &  34.29 & 33.14 & 32.57 & 0.7130\\
         \emph{CN-S}  & \textbf{56.25} & 4.17 & 39.58 & $<$0.01 \\
         \emph{CN-L} & \textbf{66.67} & 2.92 & 30.42 & $<$0.01 \\
         \hline
        \hline
    \end{tabular}
\label{tab: ABX_STRAIGHT}
\end{table}
\subsubsection{Effects of GMN}
\label{subsec3D3: Effects of GMN}

\begin{table}
\centering
    \caption{Average preference scores (\%) on speech quality between \emph{\textbf{HiNet-S}} and \emph{\textbf{HiNet-S-woGMN}} for speaker \emph{slt} when using natural acoustic features as input, where N/P stands for ``no preference" and $p$ denotes the $p$-value of a $t$-test between two vocoders.}
    \begin{tabular}{c c c c}
        \hline
        \hline
        \emph{\textbf{HiNet-S}} & \emph{\textbf{HiNet-S-woGMN}} & \emph{N/P}& $p$ \\
         \hline
          \textbf{34.09}& 22.73 & 43.18 & 0.0250\\
         \hline
        \hline
    \end{tabular}
\label{tab: ABX_GMN}
\end{table}

As introduced in Section \ref{subsec2A: Amplitude Spectrum Predictor}, GMN was introduced to compensate the global distortion between the predicted amplitude spectra and natural ones.
To confirm the effectiveness of GMN, the \emph{\textbf{HiNet-S-woGMN}} vocoder was built for comparing with \emph{\textbf{HiNet-S}}.
This vocoder was a simplified HiNet vocoder without GMN.
Table \ref{tab: ABX_GMN} listed the results of the ABX subjective test for speaker \emph{slt}.
We can see that \emph{\textbf{HiNet-S}} outperformed \emph{\textbf{HiNet-S-woGMN}} significantly ($p <$0.05).
This result shows that GMN had a positive effect on improving the performance of the HiNet vocoder without using GANs for speaker $slt$.

\subsubsection{Effects of Pre-Calculated Initial Phase}
\label{subsec3D4: Effects of Pre-Calculated Phase}
\begin{figure}[t]
    \centering
    \includegraphics[height=5.5cm]{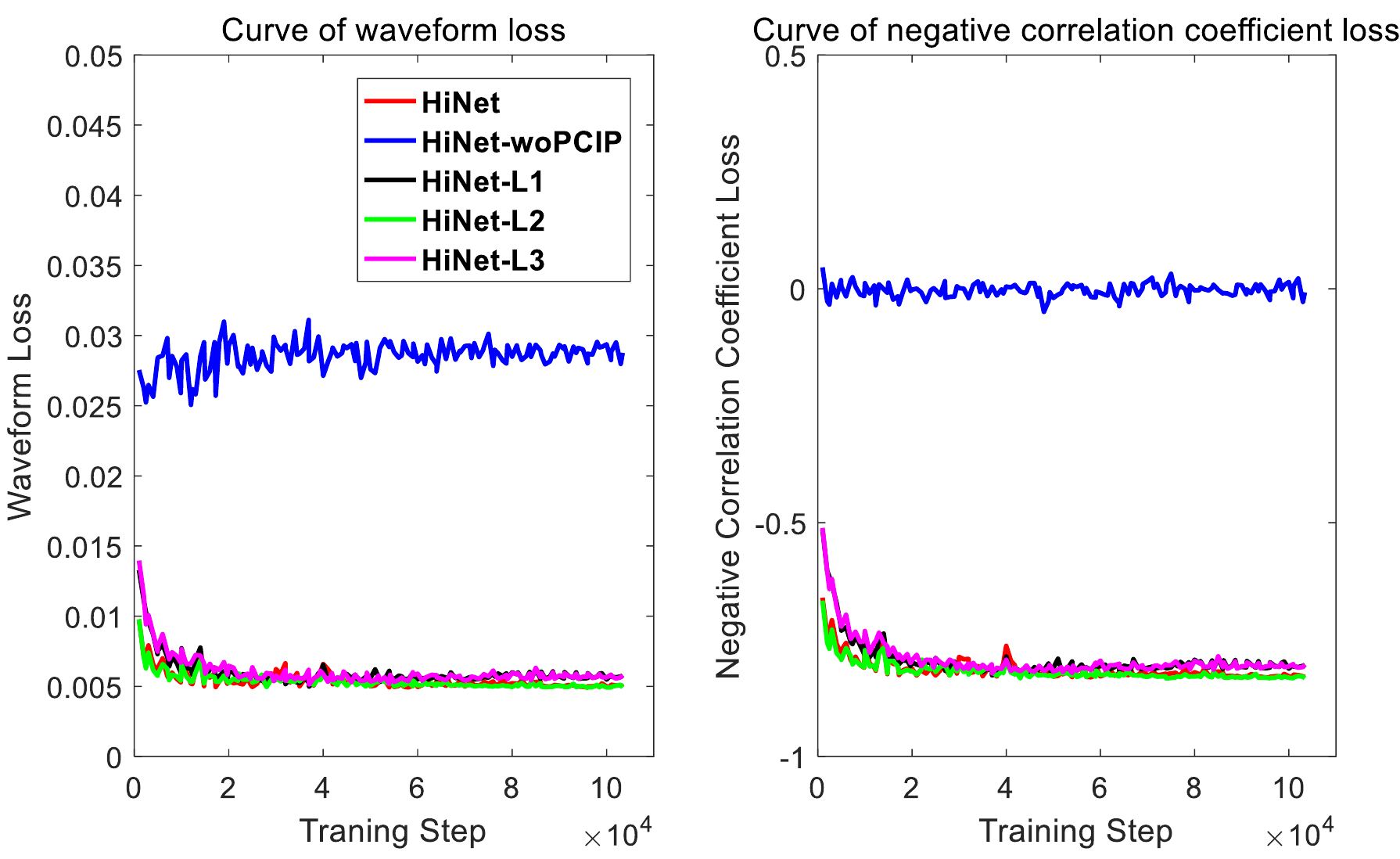}
    \caption{The waveform loss and the negative correlation coefficient loss of PSPs in different vocoders on the validation set of speaker \emph{slt},
     where the x-axis shows training steps.}
    \label{fig: Curve}
\end{figure}

\begin{table}[t]
\centering
    \caption{SNRs among the original HiNet vocoder, the HiNet vocoder without pre-calculated initial phase and the HiNet vocoders trained with different loss functions on the test sets of speaker \emph{slt} when using natural acoustic features as input.}
    \resizebox{8.8cm}{0.65cm}{
    \begin{tabular}{c | c c c c c c}
        \hline
        \hline
         & \emph{\textbf{HiNet}}& \emph{\textbf{HiNet-woPCIP}}& \emph{\textbf{HiNet-L1}} & \emph{\textbf{HiNet-L2}} & \emph{\textbf{HiNet-L3}}\\
         \hline
         SNR(dB) & \textbf{6.2937}& 1.8579 & 6.1380 & 6.2899 & 6.1377\\
         SNR-V(dB) & 8.9254 & 2.4820 & 8.6689 & \textbf{8.9394} & 8.6683\\
        \hline
        \hline
    \end{tabular}}
\label{tab_LF}
\end{table}

As introduced in Section \ref{subsec2B1: Source Module}, we pre-calculated the initial phase $\phi_j$  for the sine-based excitation signal of each voiced segment at the training stage of PSP, expecting to benefit the recovery of phase spectra.
To confirm the effectiveness of the pre-calculated initial phase,  the \emph{\textbf{HiNet-woPCIP}} vocoder was built for comparison.
This vocoder adopted random initial phase $\phi_j$ for the sine-based excitation signal of all voiced segments at the training stage of PSP which was different from \emph{\textbf{HiNet}}.
The speaker \emph{slt} was used for experiments.

Here we focused on the SNR and SNR-V metrics which reflected the performance of phase prediction and the results are listed in Table \ref{tab_LF}.
It is obvious that \emph{\textbf{HiNet-woPCIP}} achieved much lower waveform SNR and SNR-V than \emph{\textbf{HiNet}}. 
Fig. \ref{fig: Curve} draws the curves of the waveform loss and the negative correlation coefficient loss of PSPs on the validation set as a function of training steps.
In our implementation, a training step generated a truncated sequence with 16000 samples.
An epoch contained 2462 training steps for speaker \emph{slt}.
A validation was performed every 1000 training steps and at the end of an epoch during the training process.
We can see from Fig. \ref{fig: Curve} that the waveform loss and the negative correlation coefficient loss of \emph{\textbf{HiNet}} gradually decreased and both converged eventually, which implied that the PSP in the original HiNet vocoder gradually learnt the phase information during model training.
However, the waveform loss of \emph{\textbf{HiNet-woPCIP}} was almost unchanged and its negative correlation coefficient loss remained close to zero (i.e., no correlation), which indicated that discarding the pre-calculated initial phase prevented the PSP from learning the phase information through  the waveform loss and the negative correlation coefficient loss.
Therefore, the initial phase pre-calculation was crucial in our proposed method. 

\subsubsection{Effects of Loss Functions}
\label{subsec3D5: Effects of Different Loss Functions}

As introduced in Section \ref{subsec2B3: Training Criteria}, a combination of amplitude spectrum loss, waveform loss and negative correlation coefficient loss was used
to train the waveform generator in PSP.
In this subsection, we explored the effects of the components in the combined loss function by ablation tests.
Three vocoders with different loss functions for PSP compared with \emph{\textbf{HiNet}} were built and their descriptions were as follows.
\begin{itemize}
\item \emph{\textbf{HiNet-L1}} The HiNet vocoder removing the negative correlation coefficient loss from the combined loss function for PSP (i.e., $\mathcal L_{Comb}=\sum_i\mathcal L_{ASi}+\mathcal L_{W}$).
\item \emph{\textbf{HiNet-L2}} The HiNet vocoder removing the waveform loss from the combined loss function for PSP (i.e., $\mathcal L_{Comb}=\sum_i\mathcal L_{ASi}+\mathcal L_{C}$).
\item \emph{\textbf{HiNet-L3}} The HiNet vocoder removing the waveform loss and the negative correlation coefficient loss from the combined loss function for PSP (i.e., $\mathcal L_{Comb}=\sum_i\mathcal L_{ASi}$).
\end{itemize}

\begin{table}[t]
\centering
    \caption{Average preference scores (\%) on speech quality between \emph{\textbf{HiNet}} and \emph{\textbf{HiNet-L3}} of speaker \emph{slt} when using natural acoustic features as input, where N/P stands for ``no preference" and $p$ denotes the $p$-value of a $t$-test between two vocoders.}
    \begin{tabular}{c c c c}
        \hline
        \hline
        \emph{\textbf{HiNet}}& \emph{\textbf{HiNet-L3}} & \emph{N/P}& $p$ \\
         \hline
         30.00 & 24.68 & 45.32 & 0.0731\\
         \hline
        \hline
    \end{tabular}
\label{tab: ABX_LF}
\end{table}

Similar with Section \ref{subsec3D4: Effects of Pre-Calculated Phase}, only the SNR and SNR-V results on speaker \emph{slt} are listed in Table \ref{tab_LF}.
Comparing \emph{\textbf{HiNet}} with \emph{\textbf{HiNet-L1}}, it can be observed that
removing the negative correlation coefficient loss led to the degradation on waveform SNR and more significant degradation on SNR-V.
In contrast, removing the waveform loss did not cause a significant degradation on waveform SNR.
The curves of the waveform loss and the negative correlation coefficient loss on the validation set are also drawn in Fig. \ref{fig: Curve}.
We can see that there were no significant differences between \emph{\textbf{HiNet}} and \emph{\textbf{HiNet-L2}}.
However, the  converged losses of the HiNet vocoders trained without the negative correlation coefficient loss (i.e., \emph{\textbf{HiNet-L1}} and \emph{\textbf{HiNet-L3}}) were slightly higher than that of
the other two HiNet vocoders (i.e., \emph{\textbf{HiNet}} and \emph{\textbf{HiNet-L2}}).
We also conducted a subjective ABX test to compare \emph{\textbf{HiNet}} with \emph{\textbf{HiNet-L3}}.
The results are listed in Table \ref{tab: ABX_LF}.
We can see that \emph{\textbf{HiNet}} was better than \emph{\textbf{HiNet-L3}} but the difference was not obviously significant ($p$ slightly greater than 0.05).
In summary, the negative correlation coefficient loss and the waveform loss had a positive impact on objective results but had little significant impact in subjective evaluations.

\section{Conclusion}
\label{sec4: Conclusion}

In this paper, we have proposed a novel neural vocoder named HiNet which adopts hierarchical generation of amplitude and phase spectra for statistical parametric speech synthesis.
The HiNet vocoder consists of an amplitude spectrum predictor (ASP) and a phase spectrum predictor (PSP).
The former employs a DNN model to generate the amplitude spectra  and the latter utilizes a neural source-filter (NSF) waveform generator to predict the phase spectra given
amplitude spectra.
The experimental results show that our proposed HiNet vocoder outperformed the conventional STRAIGHT vocoder, a 16-bit WaveNet vocoder using open source implementation and an NSF vocoder with similar complexity to the PSP, and achieved similar performance with a 16-bit WaveRNN vocoder.
Because there are no autoregressive structures in both ASP and PSP, our proposed HiNet vocoder can reconstruct speech waveforms very efficiently.
Through model simplification, the proposed HiNet vocoder can 
generate 1s waveforms of 16kHz speech in about 0.19s.
Further improving the performance of ASP and PSP by using other advanced model structures and
applying the HiNet vocoder to other tasks such as voice conversion will be  our future work.

\bibliographystyle{IEEEtran}
\bibliography{mybib}
\end{document}